\documentclass[12pt]{article}

\pdfoutput=1 

\usepackage[T1]{fontenc}
\usepackage[utf8]{inputenc}
\usepackage[english]{babel}
\usepackage{lmodern}
\usepackage{amsmath}
\usepackage{amsthm}
\usepackage{amssymb}
\usepackage{mathrsfs}
\usepackage{bm}
\usepackage{braket}
\usepackage{bbm}
\usepackage{mathtools}
\usepackage{ifthen}
\usepackage{slashed}
\usepackage{graphicx}
\usepackage{epstopdf}       
\usepackage{xspace}
\usepackage{microtype}
\usepackage{booktabs}
\usepackage{array, tabularx}
\usepackage{chngpage}
\usepackage[english]{varioref}
\usepackage[nottoc]{tocbibind}
\usepackage[numbers,sort&compress]{natbib}
\usepackage{url}
\usepackage{multicol, multirow}
\usepackage{mparhack}
\usepackage{emptypage}
\usepackage{setspace}
\usepackage{dsfont}
\usepackage{rotating}
\usepackage{enumitem}
\usepackage{comment}
\usepackage{tikz,tikz-cd}
\usepackage{xcolor}

\usetikzlibrary{cd,shadings}

\numberwithin{equation}{section}  

\usepackage[left=2.5cm, right=2.5cm, top=2.5cm, bottom=3cm]{geometry}
\usepackage[colorlinks=true
,urlcolor=blue
,anchorcolor=blue
,citecolor=blue
,filecolor=blue
,linkcolor=blue
,menucolor=blue
,linktocpage=true
,pdfproducer=medialab
,pdfa=true
]{hyperref}

\renewcommand{\tilde}{\widetilde}

\renewcommand{\hat}{\widehat}

\renewcommand{\bar}{\overline}

\DeclareMathOperator{\Tr}{Tr}

\DeclareMathAlphabet{\mathbfsf}{OT1}{cmss}{bx}{n}

\newcommand{\Z}{\mathbb{Z}}
\newcommand{\C}{\mathbb{C}}

\newcommand{\mb}[1]{\mathbf{#1}}
\newcommand{\mc}[1]{\mathcal{#1}}

\newcommand{\mf}[1]{\mathfrak{#1}}

\newcommand{\cA}{\mathcal{A}}

\newcommand{\cD}{\mathcal{D}}

\newcommand{\cF}{\mathcal{F}}

\newcommand{\cN}{\mathcal{N}}

\newcommand{\be}{\begin{equation}}
\newcommand{\ee}{\end{equation}}
\newcommand{\beq}{\begin{equation}}
\newcommand{\eeq}{\end{equation}}

\newcommand{\ii}{{\rm i}}
\newcommand{\e}{{\rm e}}

\newcommand{\g}{\mf{g}}
\newcommand{\rd}{{\rm d}}
\newcommand{\dd}{{\rm d}}
\newcommand{\abs}[1]{\left| #1 \right|}
\newcommand{\zbar}{\overline{z}}
\newcommand{\vol}{{\rm vol}}
\newcommand{\ph}[1]{\phantom{#1}}
\newcommand{\identity}{\mathbbm{1}}

\newcommand{\mexp}{\mathtt{g}}

\newcommand{\multiplet}[4]{\left[ #1 \right]_{ #2 }^{\left( #3 , #4 \right)}}
\renewcommand{\j}{\varphi}
\newcommand{\st}{\varphi}
\newcommand{\gc}{\mathfrak{g}}

\newcommand{\ma}{\mathrm{a}}
\renewcommand{\S}{\mathbb{S}}
\newcommand{\Xsource}{\mc{X}}
\newcommand{\Xop}{\mc{O}_{\Delta_-}}
\newcommand{\jop}{\mc{O}_{\Delta_+}}
\newcommand{\tredSUGRAQ}{\zeta}
\newcommand{\tredSUGRAS}{\vartheta}
\newcommand{\E}{{\rm E}}

\allowdisplaybreaks 

\begin{document} 

\begin{titlepage}

\begin{center}
\today

\vspace{1cm}

{\LARGE \bf Boundary Conditions \\[5pt] in Topological AdS$_4$/CFT$_3$}

\vspace{1.2cm}

{Pietro Benetti Genolini$^a$, Matan Grinberg$^{a,b}$ and Paul Richmond$^c$}

\vspace{0.8cm}

{$^a${\it Department of Applied Mathematics and Theoretical Physics, \\
University of Cambridge, Wilberforce Road, Cambridge, CB3 OWA, UK}}

\vspace{0.3cm}

{$^b${\it Department of Physics, \\
Princeton University, Princeton, NJ 08544, USA}}

\vspace{0.3cm}

{$^c${\it Department of Mathematics, \\
King's College London, Strand, WC2R 2LS, UK}}

\vspace{0.8cm}

{\small \tt Pietro.BenettiGenolini@damtp.cam.ac.uk,}
\\{\small \tt matang@princeton.edu / mg993@cam.ac.uk,}
\\{\small \tt Paul.Richmond@kcl.ac.uk}

\vspace{2.5cm}

{\bf {\sc Abstract}} 
\end{center}
We revisit the construction in four-dimensional gauged $Spin(4)$ supergravity of the holographic duals to topologically twisted three-dimensional $\mathcal{N}=4$ field theories. Our focus in this paper is to highlight some subtleties related to preserving supersymmetry in AdS/CFT, namely the inclusion of finite counterterms and the necessity of a Legendre transformation to find the dual to the field theory generating functional. Studying the geometry of these supergravity solutions, we conclude that the gravitational free energy is indeed independent from the metric of the boundary, and it vanishes for any smooth solution.

\end{titlepage}

\tableofcontents

\section{Introduction}
\label{sec:Intro}

Observables of strongly-coupled field theories may be related to semi-classical gravity on spaces that are asymptotically anti-de Sitter following the AdS/CFT correspondence. In order to further sharpen the dictionary between gauge theory and gravity, it is useful to look at supersymmetric field theories that are amenable to localization, as in this case protected observables may be computed exactly. We begin by formulating a supersymmetric field theory on a Riemannian manifold $(M_d, g)$ and computing its partition function $Z[M_d]$. We then look for a gravitational dual: a solution $(Y_{d+1},G)$ to an appropriate supergravity theory which is asymptotically locally hyperbolic with conformal boundary $(M_d,g)$. That is, we require the existence of a boundary isomorphic to $M_d$ and a coordinate $z$ such that near the boundary at $z=0$, $G \sim \frac{\rd z^2}{z^2} + \frac{1}{z^2}g$. The AdS/CFT prescription, in broad outline, is that in an appropriate limit
\beq
\label{eq:AdSCFT}
Z[M_d] \ = \ \sum \e^{- \mathbb{S}[Y_{d+1}]} \, ,
\eeq
where $\mathbb{S}$ is the holographically renormalized supergravity action.

In addition to fixing the conformal class of the boundary metric, supersymmetry on a curved background generically requires additional structures, which must also be matched by the gravitational dual. For instance, focusing on $d=3$, the minimum amount of supersymmetry necessary for a localization computation is $\mc{N}=2$, which requires the background to admit a transversely holomorphic foliation, that is, to be a Seifert manifold (or torus bundles over a circle) \cite{Closset:2012ru}. 
As we increase the amount of supersymmetry of the boundary theory, we can preserve supercharges on arbitrary curved spaces using the (full) topological twist \cite{Witten:1988ze}. In this case, the partition function of the field theory is a diffeomorphism invariant, as it does not depend on the background metric, and in fact it may reproduce invariants studied elsewhere in mathematics: famously, Witten's twist of four-dimensional $\mc{N}=2$ SYM reproduces Donaldson's invariants. It is natural to ask ourselves what would the gravity dual be. The strategy employed in \cite{BenettiGenolini:2017zmu, GRS1} was to study the supersymmetry structure of the supergravity solutions with boundary conditions corresponding to the (generalized) Killing spinors and bosonic fields required to perform the topological twist of the field theory.\footnote{An analogous strategy has also been applied to the hyperbolic dual of the four-dimensional $\mc{N}=2$ $\Omega$ deformation \cite{Bobev:2019ylk, BenettiGenolini:2019wxg}.}

\medskip

The importance of studying holography in the context of topological field theories stems from the fact that in these contexts the field theory is under better control, and so it should allow us to improve our understanding of the right-hand side of \eqref{eq:AdSCFT}, which is the classical limit of the path integral for quantum gravity. For us, this means imposing stringent conditions on the domain of the sum (e.g.\ should we include real or complex solutions? Smooth or singular?). However, there have also been numerous exciting recent developments on complementary approaches to the definition of a ``topological'' subsector of supergravity, including attempts to twisting the supergravity/string theory itself (or sectors thereof) \cite{Costello:2016mgj, Bonetti:2016nma, Mezei:2017kmw, Costello:2017fbo, Costello:2018zrm, Gaiotto:2019wcc, Costello:2020jbh, Oh:2020hph, Gaiotto:2020vqj}, and twisting three-dimensional quantum gravity \cite{Li:2019qzx, Li:2020nei}. Furthermore, there have also been studies on the localization induced by supersymmetry in the form of the supergravity path integral \cite{Dabholkar:2014wpa, Bae:2015eoa, Imbimbo:2018duh, Hristov:2018lod, deWit:2018dix, Jeon:2018kec} and of the classical gravity on-shell action \cite{BenettiGenolini:2019jdz}.

\medskip

The additional control over the field theory provided by the full topological twist also allows us to better study another issue in the AdS/CFT dictionary: the relevance of boundary conditions on the gravity side. Namely, there are two issues that we will discuss.
Firstly, renormalizing the gravitational on-shell action requires a choice of regularization scheme, and choosing a scheme that is not compatible with supersymmetry creates numerous puzzling issues with the holographic dictionary \cite{FreedmanPufu, FreedmanPufuEtAl, Genolini:2016ecx, Halmagyi:2017hmw, Cabo-Bizet:2017xdr}. Secondly, the extended supersymmetry necessary to perform the twist requires the presence of scalars in the bulk. Scalars fields in gauged supergravities may have different boundary conditions subtly dependent on their masses and the spacetime dimension \cite{Breitenlohner:1982jf}.\footnote{\textit{A priori} other fields could also have multiple boundary conditions \cite{Breitenlohner:1982jf, Mezei:2017kmw}, the only relevant ones in this paper are the scalars.} A scalar with mass $m$ on an asymptotically locally AdS space of dimension $d+1$ admits a unique boundary condition compatible with the symmetries of AdS$_{d+1}$ if
\beq
\label{eq:BF1}
m^2 > - \frac{d^2}{4}+1 \, ,
\eeq
but may have two different boundary conditions (and thus two quantizations) if
\beq
\label{eq:BF2}
- \frac{d^2}{4} < m^2 < - \frac{d^2}{4}+1 \, .
\eeq
In the context of the AdS/CFT correspondence, the scaling dimension $\Delta$ of a field theory operator $\mc{O}_{\Delta}$ dual to a bulk scalar with dimension $m$ is given by 
\beq
\label{eq:AdSCFTScaling}
\Delta_{\pm} \ = \ \frac{d \pm \sqrt{d^2 + 4m^2}}{2} \, ,
\eeq
and the allowed scaling dimensions should satisfy the unitarity bound $\Delta > (d-2)/2$. If \eqref{eq:BF1} holds, then only $\Delta_+$ satisfies the unitarity bound and is allowed,
so we can only have an operator $\mc{O}_{\Delta_+}$. Near the boundary at $z=0$, the leading term in the expansion of the corresponding scalar is the coefficient of $z^{d-\Delta_+}$, which we interpret as the source for $\mc{O}_{\Delta_+}$, and we set as a boundary condition for the Dirichlet problem in supergravity. However, if instead we are in the regime \eqref{eq:BF2}, both $\Delta_{\pm}$ are allowed scaling dimensions for the boundary operators, corresponding to the two quantizations allowed for the bulk scalar, and the possibility of $\mc{O}_{\Delta_-}$ requires additional care. In this case, the source of the boundary $\mc{O}_{\Delta_-}$ is not the leading term in the expansion of the bulk scalar near the boundary,  and the generating functional of the field theory cannot be identified with the renormalized gravitational on-shell action, since we cannot set the appropriate boundary conditions for the Dirichlet problem. In fact, the generating functional in this case is given by the Legendre transform of the renormalized on-shell action \cite{Klebanov:1999tb}.

\medskip

Whilst it is true that in the range \eqref{eq:BF2} there are two possible quantizations of the bulk scalar fields, and correspondingly boundary operators with two different scaling dimensions belonging to two different field theories with generating functional related by Legendre transformation, it is not obvious that both of them are supersymmetric. For the $\mc{N}=4$ gauged supergravity relevant to this paper, $d=3$ and there are two scalars with $m^2 = -2$ \cite{DFR}. Since the mass of the scalars satisfies \eqref{eq:BF2}, there are two potential quantizations, but only one of them is compatible with the bulk supersymmetry, namely one including \textit{both} boundary conditions set by $\Delta_+ =2$ and $\Delta_- = 1$, sometimes referred to as alternate boundary conditions \cite{Breitenlohner:1982jf}. Correspondingly, the supermultiplet of the stress-energy tensor of three-dimensional local superconformal field theories with $\mc{N} = 4$ contains two scalar operators $\mc{O}_{\Delta_+}, \mc{O}_{\Delta_-}$ with scaling dimensions $2$ and $1$ respectively.

\medskip

In this paper we will study the gravity dual to the full topological twist of three-dimensional $\mc{N}=4$ superconformal field theories, highlighting the importance of boundary conditions preserving supersymmetry and studying the subtleties explained above. In particular, we will show that we can only obtain a gravitational observable dual to the field theory generating functional that is independent of the boundary metric if we include supersymmetry-preserving counterterms in the holographic renormalization \textit{and} if we perform the appropriate Legendre transformation on the renormalized on-shell action. Furthermore, we will evaluate the on-shell action for smooth supergravity solutions, proving that it is only by including supersymmetric counterterms and Legendre transforming that we obtain a topological invariant of the boundary.

\medskip

The same supergravity had been considered in \cite{GRS1}, but with both scalars being dual to boundary operators with scaling dimension $\Delta_+ = 2$. Several puzzling features were alluded to there but not fully understood. Here, we investigate in more detail the subtleties of the holographic renormalization in the presence of alternate boundary conditions.
However, the results of this paper do not modify the conclusions of \cite{GRS1}. We also show that a supersymmetry-preserving holographic renormalization scheme, with twisted boundary conditions, results in the gravitational observable appearing in \eqref{eq:AdSCFT} being zero for all smooth supergravity solutions. Under the assumption that smooth real solutions dominate the gravity saddle approximation, this leads to a conjectural behaviour of the large $N$ limit of the partition function for the topologically twisted ABJM theory on any Riemannian three-manifold: it should be $o(N^{3/2})$.\footnote{Note that any spin $M_3$ bounds a smooth $Y_4$, so a smooth gravitational filling to the Dirichlet problem always exists.} Apart from the Vafa--Witten twisted theory on $K3$ \cite{Vafa:1994tf}, no large $N$ limit of a (fully) twisted partition function is known. If the conjecture for the behaviour of the field theory were not to hold, we should infer that \textit{singular} or \textit{complex} solutions dominate the quantum gravity path integral. 

\newpage
\noindent
\textbf{Outline}

\medskip

\noindent
In Section \ref{sec:FieldTheory} we introduce the field theory side: first, we review from an abstract viewpoint the topological twists of three-dimensional field theories with eight real supercharges, then we describe the coupling of the superconformal field theories to off-shell conformal supergravity and in particular the solution of the resulting Killing spinor equations on an arbitrary three-manifold. In Section \ref{sec:DualSUGRA} we move to the bulk supergravity theory, reviewing its supersymmetry transformations and its equations of motion, and the results of their Fefferman--Graham expansion. We then describe in some detail how to apply holographic renormalization to this theory without spoiling supersymmetry. Finally, in Section \ref{sec:BulkTwist} we look at the on-shell gravitational action when we fix twisted boundary conditions, showing that it is independent of the boundary metric and it evaluates to a topological invariant. In order to do so, we will highlight the importance of preserving supersymmetry, both in the renormalization scheme and in the necessity of the Legendre transformation. We also include three appendices, describing the reduction from maximal $\mc{N}=8$ gauged supergravity to the $\mc{N}=4$ model, and some considerations on the global supersymmetry of the supergravity action.

\section{Field theory}
\label{sec:FieldTheory}

We begin this section by succinctly reviewing some aspects of the topological twists in three dimensions. We will then describe how to obtain a twisted $\mc{N}=4$ superconformal field theory by coupling to off-shell conformal supergravity.

\subsection{Topological twists in three dimensions}
\label{subsec:TQFT}

In the approach to rigid supersymmetry of \cite{Festuccia:2011ws}, we preserve supersymmetry on a curved space by coupling a field theory supersymmetric on flat space to an off-shell formulation of supergravity, then decoupling the dynamics of gravity while choosing a background configuration corresponding to the required manifold. Concretely, this requires solving the (generalized) Killing spinor equations obtained from the vanishing of the supersymmetry variations of the spinors in the gravity supermultiplet.

Performing a topological twist on a field theory means in particular being able to preserve a supersymmetric charge on an arbitrary Riemannian manifold. Therefore, it corresponds to finding a configuration of the bosonic background fields in the gravity supermultiplet such that the Killing spinor equation admits a solution on any manifold. Geometrically, supersymmetry provides us with an $R$-symmetry gauge bundle $P_R$ with associated vector bundle $V_R$ and connection $A^R$. On the other hand, we have an $SO(d)$ structure on the Riemannian background. Twisting corresponds to finding a vector bundle $V$ associated to the $SO(d)$ structure such that we can identify $V$ with $V_R$ and $\omega$, the connection on $V$, with $A^R$. This identification requires a sufficiently large $R$-symmetry group and a correspondingly large amount of supersymmetry. In particular, in three dimensions it requires at least eight real supercharges, or $\mc{N}=4$ in terms of the minimal spinor.

\medskip

The $SO(3)$ structure on any oriented Riemannian three-manifold can be lifted to a $Spin(3)_E \cong SU(2)_E$ structure, as any oriented three-dimensional manifold is spin. On the other hand, the three-dimensional $\mc{N}=4$ superalgebra has a $Spin(4)_R \cong SU(2)_+\times SU(2)_-$ automorphism group. Thus, we see that there are at least two obvious twists, depending on which $SU(2)$ $R$-symmetry subbundle we choose to identify with the $SU(2)_E$ tangent bundle. Even though in the algebra there is an additional $\Z_2$ automorphism exchanging $SU(2)_+$ and $SU(2)_-$, the two twists are generically not equivalent, as the two $SU(2)$ factors are subtly different. For instance, if we construct $3d$ $\mc{N}=4$ SYM by reducing six-dimensional $\mc{N}=(1,0)$, the difference between the two subgroups arises because $SU(2)_-$ represents the $R$-symmetry group of the higher-dimensional theory, whereas $SU(2)_+$ represents the rotations in the reduced three dimensions.
More concretely, we may already see the difference in the twists by considering the transformation of the vector multiplet. The three-dimensional $\mc{N}=4$ vector multiplet comprises the gauge vector field $A_i$, gaugino $\lambda_\alpha$ and a triplet of real scalars $\vec{\phi}$, all valued in the gauge algebra: under the group $SU(2)_E\times SU(2)_+\times SU(2)_-$, they transform as
\beq
\label{eq:N4vm}
A_i \quad \multiplet{\mathbf{3}}{}{\mathbf{1}}{\mathbf{1}} \, , \qquad \lambda_\alpha \quad \multiplet{\mathbf{2}}{}{\mathbf{2}}{\mathbf{2}} \, , \qquad \vec{\phi} \quad \multiplet{\mathbf{1}}{}{\mathbf{3}}{\mathbf{1}} \, ,
\eeq
where the label between square brackets is the dimension of the $SU(2)_E$ representation, and in the superscript are the dimensions of the representations under $SU(2)_+\times SU(2)_-$.
The scalars transform in the adjoint of $SU(2)_+$ but are singlets under $SU(2)_-$ (which are often labelled $SU(2)_C$ and $SU(2)_H$ to highlight their action on the Coulomb and Higgs branches), so twisting with one or the other will result in a different field content. Namely, twisting with $SU(2)_-$ leaves a symmetry group $\left( SU(2)_E\times SU(2)_- \right)_{\rm diag}\times SU(2)_+$, under which the fields transform as
\beq
\multiplet{\mathbf{3}}{}{\mathbf{1}}{\mathbf{1}} \ \to \ \left[ \mathbf{3} \right]^{\mathbf{1}} \, , \qquad \multiplet{\mathbf{2}}{}{\mathbf{2}}{\mathbf{2}} \to \left[\mathbf{1}\right]^{\mathbf{2}} \oplus \left[ \mathbf{3} \right]^{\mathbf{2}} \, , \qquad \multiplet{\mathbf{1}}{}{\mathbf{3}}{\mathbf{1}} \to \left[ \mathbf{1} \right]^{\mathbf{3}} \, .
\eeq
This twist, referred to as the A-twist, is the reduction of the four-dimensional Witten twist \cite{Witten:1989sx}. Alternatively, one can twist with $SU(2)_+$, leaving a symmetry group $\left( SU(2)_E\times SU(2)_+ \right)_{\rm diag}\times SU(2)_-$, under which the fields transform as
\beq
\multiplet{\mathbf{3}}{}{\mathbf{1}}{\mathbf{1}} \ \to \ \left[ \mathbf{3} \right]^{\mathbf{1}} \, , \qquad \multiplet{\mathbf{2}}{}{\mathbf{2}}{\mathbf{2}} \to \left[\mathbf{1}\right]^{\mathbf{2}} \otimes \left[ \mathbf{3} \right]^{\mathbf{2}} \, , \qquad \multiplet{\mathbf{1}}{}{\mathbf{3}}{\mathbf{1}} \to \left[ \mathbf{3} \right]^{\mathbf{1}} \, .
\eeq
This second twist, the B-twist, is instead intrinsically three-dimensional \cite{Blau:1991bn}. 

Similarly, the difference between the two twists can be seen by applying them to the other $\mc{N}=4$ supermultiplet, the hypermultiplet, containing two complex scalars $q$ and two spinors $\psi$ transforming as
\beq
q \quad \multiplet{\mathbf{1}}{}{\mathbf{1}}{\mathbf{2}} \, , \qquad \psi_\alpha \quad \multiplet{\mathbf{2}}{}{\mathbf{2}}{\mathbf{1}} \, .
\eeq
Parallel to both vector multiplets and hypermultiplets, there are also twisted vector multiplets and hypermultiplets, where the r\^oles of $SU(2)_+$ and $SU(2)_-$ are exchanged (the name is not connected to the topological twist).

In three dimensions, in addition to Yang--Mills theory, it is possible to preserve $\mc{N}=4$ supersymmetry in the presence of matter when the gauge field has a Chern--Simons interaction \cite{Gaiotto:2008sd}, and it is also possible to twist the resulting theory \cite{Kapustin:2009cd, Koh:2009um}, obtaining A, B and AB-twists (the latter only if the theory contains the same number of hypermultiplets and twisted hypermultiplets).

\medskip

In addition to the property of preserving supersymmetry on an arbitrary Riemannian manifold, topologically twisted theories have the property that the supersymmetry-protected partition function and other BPS observables do not depend on the background metric. In the case of the four-dimensional Witten twist, this famously allows one to recover the Donaldson invariants of the manifold \cite{Witten:1988ze}, and its three-dimensional reduction, the A-twist of $3d$ SYM, gives the Casson (or Casson--Lescop--Walker) invariant \cite{Witten:1989sx}, and via RG flow the Rozansky--Witten invariant \cite{Mikhaylov:2015nsa}. It is clear that the protected observables computed in the twisted Chern--Simons-matter models introduced above should also correspond to topological invariants of the manifold. However, their mathematical content is not yet fully clear.

\subsection{Off-shell conformal supergravity and the topological twist}
\label{subsec:BdryOffShell}

With our choice of supergravity, we will describe features of the dynamics of the stress-energy tensor multiplet of superconformal field theories (with holographic duals), which we now describe. The supermultiplet containing the stress-energy tensor is composed of \cite{Cordova:2016emh}
\begin{equation}
\label{eq:3dTmultiplet}
\begin{tabular}{cccccccc}
\toprule
Field & $\mc{O}_{\Delta_-}$ & $\lambda^a_\alpha$ & $\mathscr{J}^I_i$ & $\hat{\mathscr{J}}^{I}_i$ & $\mc{O}_{\Delta_+}$ & $\mc{S}_{i \alpha}^a$ & $T_{ij}$ \\[7pt]
{\small $\multiplet{SU(2)_E}{U(1)}{SU(2)_+}{SU(2)_-}$} & $\multiplet{\mathbf{1}}{1}{\mathbf{1}}{\mathbf{1}}$ & $ \multiplet{\mathbf{2}}{\frac{3}{2}}{\mathbf{2}}{\mathbf{2}} $ & $\multiplet{\mathbf{3}}{2}{\mathbf{3}}{\mathbf{1}}$ & $\multiplet{\mathbf{3}}{2}{\mathbf{1}}{\mathbf{3}}$ & $\multiplet{\mathbf{1}}{2}{\mathbf{1}}{\mathbf{1}}$ & $ \multiplet{\mathbf{4}}{\frac{5}{2}}{\mathbf{2}}{\mathbf{2}}$ & $\multiplet{\mathbf{5}}{3}{\mathbf{1}}{\mathbf{1}} $ \\
\bottomrule
\end{tabular}
\end{equation}

\noindent
Here the labels are the same as in \eqref{eq:N4vm}, with the addition of the scaling dimension written as a subscript, thus forming the full bosonic subalgebra of the $\mc{N}=4$ superconformal algebra. The indices $i,j/\alpha$ are vector/spinor spacetime indices, $a = 1, \dots, 4$ labels the component of a vector in the fundamental of $Spin(4)_R$, and $I=1, 2, 3$ is an index for the adjoint of $SU(2)$. We see that there are two scalars with different scaling dimension, the $Spin(4)_R$ $R$-symmetry current composed of two $SU(2)$ currents, each transforming in the adjoint of one of the $SU(2)$ subgroups, the supercurrent, and the stress-energy tensor.

\medskip

The natural off-shell supergravity that couples to the stress-energy tensor supermultiplet of a SCFT is conformal supergravity. In order to define rigid supersymmetric curved backgrounds for a theory, it is often preferable to use a non-conformal supergravity, even if the theory is conformally invariant on flat space, because of the ultraviolet regularization (see for instance \cite{Dumitrescu:2016ltq}). However, it is conformal supergravity that appears as we take the limit of gauged supergravity near the boundary of an asymptotically locally AdS solution. Therefore, we review here an off-shell formulation of $\mc{N}=4$ conformal supergravity \cite{Banerjee:2015uee}, which we will reproduce from the bulk.

Three-dimensional $\mc{N}=4$ conformal supergravity has a Weyl multiplet consisting of the fields\footnote{Here we have gauge fixed to zero the gauge field for the dilatations $b_\mu$.} 

\begin{equation}
\label{eq:3dWeylmultiplet}
\begin{tabular}{cccccccc}
\toprule
Field & $S_2$ & $(\chi^a_\alpha)^{3d}$ & $S_1$ & ${A}^I_i$ & $\hat{A}^I_i$ & $(\psi^a_{i\alpha})^{3d}$ & $g_{ij}$ \\[3pt]
Weyl weight & $-2$ & $-\frac{3}{2}$ & $-1$ & $0$ & $0$ & $\frac{1}{2}$ & $2$ \\
\bottomrule
\end{tabular}
\end{equation}
\medskip

\noindent
These represent the metric, gravitino, gauge fields for the $Spin(4)$ gauge group and auxiliary fields. We couple the Weyl multiplet to the stress-energy tensor supermultiplet using the Weyl weights assignment $w_{\Phi}=r_{\Phi}-\Delta_{\Phi}$ for the fields in the stress-energy tensor supermultiplet, where $r_{\Phi}$ is the tensorial rank of the field. Thus, it is clear that, at least at the linearized level, we preserve invariance under global Weyl transformations.

The spinor parameters for the $\mc{Q}$ and $\mc{S}$ supersymmetries are $\tredSUGRAQ^a$ and $\tredSUGRAS^a$, both transforming in the $\mb{4}$ of $Spin(4)_R$.\footnote{The authors of \cite{Banerjee:2015uee} use the van der Waerden notation, writing the $\mb{4}$ of $Spin(4)$ as $(\mb{2},\mb{2})$. To transform between the two we use the standard invariant symbols $\sigma_a^{A\dot{A}} = (\sigma^i, \ii \identity_2)$ and $\overline{\sigma}_{a\dot{A}A} = (\sigma^i, - \ii \identity_2)$. The left and right chiral representations are generated by
\[
\begin{split}
(S^L_{ab})^A_{\ph{A}B} \ &= \ \tfrac{1}{2}\left( \sigma_a\overline{\sigma}_b - \sigma_b \overline{\sigma}_a \right)^A_{\ph{A}B} \, , \qquad
(S^R_{ab})_{\dot{A}}^{\ph{A}\dot{B}} \ = \ \tfrac{1}{2}\left( \overline{\sigma}_a{\sigma}_b - \overline{\sigma}_b {\sigma}_a \right)_{\dot{A}}^{\ph{A}\dot{B}} \, .
\end{split}
\]
The bases of $\mf{su}(2)$ are then related by the 't Hooft symbols
\[
(\sigma^I)^A_{\ph{A}B} \ = \ - \tfrac{\ii}{4}\overline{\eta}^I_{ab}(S^L_{ab})^A_{\ph{A}B} \, , \qquad 
(\sigma^I)_{\dot{A}}^{\ph{A}\dot{B}} \ = \ - \tfrac{\ii}{4}{\eta}^I_{ab}(S^R_{ab})_{\dot{A}}^{\ph{A}\dot{B}} \, .
\]
Using these notions, it's possible to transform the equations (3.1) of \cite{Banerjee:2015uee} into our notations (with an appropriate rescaling of the gauge fields by $\sqrt{2}$ and redefinitions of the spinors).} In order to have a rigid supersymmetric background, we need to solve the (generalized) Killing spinor equations coming from setting the variations of the gravitino and the auxiliary spinor to zero. After Wick rotation, these are
\begin{align}
\label{eq:3dSUGRAGravitino}
0 \ &= \  \delta( \psi^a_i)^{3d} \ = \ \nabla_i \tredSUGRAQ^a + \tfrac{1}{2\sqrt{2}}\eta^I_{ab} A_i^I \tredSUGRAQ^b - \tfrac{1}{2\sqrt{2}}\overline{\eta}^I_{ab} \hat{A}^I_i  \tredSUGRAQ^b + \sigma_i \, \tredSUGRAS^a \, , \\
\label{eq:3dSUGRAConstraint}
0 \ &= \ \delta(\chi^a)^{3d} \ = \ \partial_i S_1 \, \sigma^i \tredSUGRAQ^a + \tfrac{1}{2} S_2 \tredSUGRAQ^a + \tfrac{1}{4\sqrt{2}} \left( \eta^I_{ab} F_{ij}^I  + \overline{\eta}^I_{ab} \hat{F}_{ij}^I \right) \sigma^{ij}\tredSUGRAQ^a - 2 S_1 \tredSUGRAS^a \, ,
\end{align}
where $\sigma^i$ are the Pauli matrices generating Cliff$(3,0)$, and the symbols $\eta^I_{ab}, \overline{\eta}^I_{ab}$ are the self-dual and anti-self-dual 't Hooft matrices respectively.
In the spirit of \cite{Festuccia:2011ws}, a background preserving $\mc{N}=4$ rigid supersymmetry in three dimensions admits spinors $\tredSUGRAQ^a, \tredSUGRAS^a$ satisfying \eqref{eq:3dSUGRAGravitino} and \eqref{eq:3dSUGRAConstraint}. There is one obvious way of solving them on arbitrary manifold, corresponding to the topological twists described earlier. We set $\hat{A}^I_i$ to zero, and then identify the remaining $SU(2)$ bundle with the tangent $Spin(3)_E$ bundle, relating the connections on the two as
\beq
\label{eq:TopTwistConnection}
A^I_i \ = \ \frac{1}{\sqrt{2}}\epsilon^I_{\ph{I}\overline{jk}}\, \omega_i^{\ph{i}\overline{jk}} \, ,
\eeq
where the overline represents frame indices and $\omega_i^{\ph{i}\overline{jk}}$ is the spin connection for the frame. It is then possible to find a spinor satisfying $\mc{D}_i\tredSUGRAQ^a = 0$ on an arbitrary manifold: in our basis it is given by
\beq
\tredSUGRAQ^a \ = \ \ii \overline{\sigma}^a \begin{pmatrix}
w \\ \ii \overline{w}
\end{pmatrix} \, ,
\eeq
where $w$ is any complex number. The remaining fields are also immediately set by the twist and the choice of spinor
\beq
\label{eq:TopTwistSUGRA}
\tredSUGRAS^a \ = \ 0 \, , \qquad S_2 \ = \ - \frac{1}{2}R \, , \qquad S_1 \ = \ 0 \, ,
\eeq
where $R$ is the Ricci scalar of the metric $g$.
In fact, the topological twist is consistent with $S_1$ being an arbitrary constant, which we set to zero. We will reproduce these conditions from the bulk in Section \ref{subsec:ExpansionSUSY}, clearing up a puzzle from \cite{GRS1}.

\subsection{Topological AdS/CFT for ABJM}
\label{subsec:ABJM}

In the following we shall discuss the gravity dual of any topologically twisted $\mc{N}=4$ SCFT (provided they admit such a dual). In order to focus on a particular field theory, we should embed our computations in ten or eleven-dimensional supergravity by choosing an uplift of the gravity solution. However, any solution of the supergravity theory we consider may be uplifted on  $S^7/\Z_k$ to a solution of eleven-dimensional supergravity \cite{Trunc}. This concretely means that we are describing the dual to topologically twisted ABJM on an arbitrary manifold.\footnote{Topological twists of the BLG model have appeared in \cite{Lee:2008cr}.}

The ABJM theory \cite{Aharony:2008ug} is a Chern--Simons-matter theory with gauge group $U(N)_k\times U(N)_{-k}$, together with hypermultiplet and twisted hypermultiplet in the bifundamental representation (in the $\mc{N}=4$ notation described in \cite{Hosomichi:2008jb}). Generically, it has $\mc{N}=6$ supersymmetry, but for $k=1,2$, this is enhanced to $\mc{N}=8$. For any $N$, it describes the infrared dynamics on the worldvolume of $N$ $M2$ branes at a $\C^4/\Z_k$ singularity. Most importantly, it is superconformal.

In the limit where $N\gg k^5$, the ABJM theory on flat space has a dual description in terms of eleven-dimensional supergravity on AdS$_4\times S^7/\Z_k$ with $N$ units of flux through the $S^7$. However, when the theory is defined on an arbitrary Riemannian three-manifold $(M_3,g)$, it becomes cumbersome to study the dual in eleven-dimensional supergravity, so we truncate eleven-dimensional supergravity on $S^7$ down to a four-dimensional gauged supergravity. In particular, in light of the discussion in the previous section, we consider the minimal supergravity necessary for the twist, which is the $\mc{N}=4$ four-dimensional supergravity with gauge group $Spin(4)$ constructed in \cite{DFR}. Since the truncation on $S^7$ from eleven to four dimensions is consistent, any solution of the $Spin(4)$ supergravity uplifts on $S^7/\Z_k$ to a solution to the eleven-dimensional equations of motion \cite{Trunc}. Therefore, any asymptotically locally AdS solution $(Y_4,G)$ uplifts to a gravity dual to the large $N$ limit of ABJM theory on $(M_3,g)$, where $(M_3,g)$ is a representative of the conformal boundary of $(Y_4,G)$, and the four-dimensional Newton constant is related to the field theory data by
\beq
\frac{1}{2\kappa^2_4} \ = \ \frac{k^{1/2}}{12\sqrt{2}\pi}N^{3/2} \, .
\eeq

\section{Dual supergravity theory}
\label{sec:DualSUGRA}

In this section we describe the relevant $\mc{N}=4$ $Spin(4)$ supergravity, then review the Fefferman--Graham expansion of the equations of motion, and finally discuss the holographic renormalization of the theory, including subtleties related to the presence of scalars.

\subsection{Action}
\label{subsec:Action}

The supergravity we consider is the four-dimensional $\mc{N}=4$ gauged supergravity with gauge group $Spin(4)$ constructed by Das--Fischler--Ro\v{c}ek \cite{DFR}. The fields in the bosonic sector are the metric $G_{\mu\nu}$, a scalar $\phi$ and a pseudoscalar $\j$, and two $SU(2)$ gauge fields $\cA^I_\mu, \hat{\cA}^I_\mu$ with associated field strengths
\beq
\cF^I \ = \ \rd \cA^I + \tfrac{1}{2}\g \, \epsilon^{IJK}\cA^J\wedge \cA^K \, , \qquad \hat{\cF}^I \ = \ \rd \hat{\cA}^I + \tfrac{1}{2}\g \, \epsilon^{IJK}\hat{\cA}^J\wedge \hat{\cA}^K \, .
\eeq
Our main focus will be the bosonic part of the action in Euclidean signature:
\beq
\label{eq:IEuclid}
\begin{split}
	I \ &= \ - \frac{1}{2 \kappa_4^2} \int \big[ R *1 - 2 X^{-2} \dd X \wedge * \dd X - \tfrac{1}{2} X^4 \dd \st \wedge * \dd \st + \gc^2 ( 8 + 2 X^2 + 2 \tilde{X}^2 ) * 1 \\
	& \qquad \quad  - \tfrac{1}{2} X^{-2} \big( \cF^I \wedge * \cF^I + \ii \st X^2 \cF^I \wedge \cF^I \big) - \tfrac{1}{2} \tilde{X}^{-2} \big( \hat{\cF}^I \wedge * \hat{\cF}^I - \ii \st X^2 \hat{\cF}^I \wedge \hat{\cF}^I \big) \big] \, ,
\end{split}
\eeq
where
\beq
X \ \equiv \ \e^{\phi/2} \, , \qquad \tilde{X}^2 \ \equiv \ X^{-2} + \j^2 X^2 \, .
\eeq
The equations of motion derived from this action are
\begin{align}
	\label{Xeom}
	\begin{split}
	0 \ =& \ \dd ( X^{-1} * \dd X ) - \tfrac{1}{2} X^4 \dd \st \wedge * \dd \st + \gc^2 \big( X^2 - X^{-2} ( 1 - \st^2 X^4 ) \big) *1  \\
	&\ + \tfrac{1}{4} X^{-2} \cF^I \wedge * \cF^I - \tfrac{1}{4} X^2 ( 1 - \st^2 X^4 ) q^{-4} \hat{\cF}^I \wedge * \hat{\cF}^I + \tfrac{\ii}{2} \st \tilde{X}^{-4} \hat{\cF}^I \wedge \hat{\cF}^I \, ,
	\end{split}\\[5pt]
	\label{AxionEOM}
	\begin{split}
	0 \ =& \ \dd ( X^4 * \dd \st ) + 4 \gc^2 X^2 \st *1 - \tfrac{\ii}{2} \cF^I \wedge \cF^I  \\
	& \ + \st X^2 \tilde{X}^{-4} \hat{\cF}^I \wedge * \hat{\cF}^I + \tfrac{\ii}{2} ( 1 - \st^2 X^4 ) \tilde{X}^{-4} \hat{\cF}^I \wedge \hat{\cF}^I \, , 
	\end{split} \\[5pt]	
	0 \ =& \ D ( X^{-2} * \cF^I ) + \ii \dd \st \wedge \cF^I \, , \label{AIeom} \\[5pt]
	0 \ =& \ \hat{D} ( \tilde{X}^{-2} * \hat{\cF}^I ) - \ii \dd ( \st X^2 \tilde{X}^{-2} ) \wedge \hat{\cF}^I \, , \label{hatAIeom} \\[5pt]	
	\label{geom}
	\begin{split}
	0 \ =& \ R_{\mu\nu} + \gc^2 G_{\mu\nu} ( 4 + X^2 + \tilde{X}^2 ) - 2 X^{-2} \partial_\mu X \partial_\nu X - \tfrac{1}{2} X^4 \partial_\mu \st \partial_\nu \st \\
	&\ - \tfrac{1}{2} X^{-2} \big( \cF^I_{\mu\rho} \cF^I_{\nu}{}^\rho - \tfrac{1}{4} G_{\mu\nu} ( \cF^I )^2 \big) - \tfrac{1}{2} \tilde{X}^{-2} \big( \hat{\cF}^I_{\mu\rho} \hat{\cF}^I_{\nu}{}^\rho - \tfrac{1}{4} G_{\mu\nu} ( \hat{\cF}^I )^2 \big)  \, ,
	\end{split}
\end{align}
where
\beq
D \cF^I \ \equiv \ \rd \cF^I + \g \epsilon^{IJK} \cA^J \wedge \cF^K \, , \qquad \hat{D} \hat{\cF}^I \ \equiv \ \rd \hat{\cF}^I + \g \epsilon^{IJK} \hat{\cA}^J \wedge \hat{\cF}^K \, .
\eeq
This theory can be obtained by truncating the four-dimensional maximal $\mc{N}=8$ gauged supergravity, as we describe in Appendix \ref{app:Reduction}. In the same appendix, we also elaborate on the supersymmetry of the theory. For the purposes of the bulk of the paper, it is only necessary to know that for supersymmetric solutions we can construct a Dirac spinor $\epsilon^a$, transforming in the fundamental of $Spin(4)$, satisfying the following equations
\begin{align}
	\label{eq:BulkGravitino}
	\begin{split}
	0 \ =& \ \mathcal{D}_\mu \epsilon^a - \tfrac{1}{8\sqrt{2}} \eta^I_{ab} X^{-1} \cF^I_{\nu\lambda} \Gamma^{\nu\lambda} \Gamma_\mu \epsilon^b + \tfrac{1}{8\sqrt{2}} \bar{\eta}^I_{ab} X^{-1} \tilde{X}^{-2} \hat{\cF}^I_{\nu\lambda} \Gamma^{\nu\lambda} \Gamma_\mu \big[ 1 + \ii \j X^2 \Gamma_5 \big] \epsilon^b \\
	& \ + \tfrac{\ii}{4} X^2 \partial_\mu \j \Gamma_5 \epsilon^a - \tfrac{1}{2\sqrt{2}} \g \big[ ( X + X^{-1} )  - \ii \j X \Gamma_5 \big] \Gamma_\mu \epsilon^a \, , 
	\end{split}\\[5pt]
	\label{eq:BulkDilatino}
	\begin{split}
	0 \ =& \ \tfrac{1}{8} \eta^I_{ab} X^{-1} \cF^I_{\nu\lambda} \Gamma^{\nu\lambda} \epsilon^b + \tfrac{1}{8} \bar{\eta}^I_{ab} X^{-1} \tilde{X}^{-2} \hat{\cF}^I_{\nu\lambda} \big[ 1 - \ii \j X^2 \Gamma_5 \big] \Gamma^{\nu\lambda} \epsilon^b \\
	&\ + \tfrac{1}{\sqrt{2}} \big[ X^{-1} \partial_\nu X + \tfrac{\ii}{2} X^2 \partial_\nu \j \Gamma_5 \big] \Gamma^\nu \epsilon^a + \tfrac{1}{2} \g \big[ ( X - X^{-1} ) + \ii \j X \Gamma_5 \big] \epsilon^a \, .
	\end{split}
\end{align}
Here, $\Gamma^\mu$ generate Cliff$(4,0)$, $\Gamma_5 \equiv - \Gamma_{1234}$, and the covariant derivative on the spinors is
\beq
\mathcal{D}_\mu \epsilon^a \ = \ \nabla_\mu \epsilon^a - \tfrac{1}{2} \g \eta^I_{ab} A^I_\mu \epsilon^b + \tfrac{1}{2} \g \bar{\eta}^I_{ab} \hat{A}^I_\mu \epsilon^b \, .
\eeq
There is a supersymmetric $\mathbb{H}^4$ vacuum solution with vanishing gauge fields and scalars $\phi = \j = 0$. Expanding the scalar terms around this vacuum, we realise that both the scalars have mass $m^2 = -4\g^2$. This will be crucial in the following analysis.

\subsection{Fefferman--Graham expansion}
\label{subsec:FGExpansion}

We now restrict our attention to asymptotically locally AdS solutions to the supergravity theory and review the Fefferman--Graham expansion of the bosonic fields. We will be brief --- the interested reader may find the analysis done in full generality in \cite{GRS1}.

In a neighbourhood of the conformal boundary, an asymptotically locally AdS metric can be written as \cite{Fefferman:2007rka}
\begin{align}
	G_{\mu\nu}\rd x^\mu \rd x^\nu \ = \ \frac{1}{z^2} \dd z^2 + \frac{1}{z^2} \mexp_{ij} \dd x^i\dd x^j \ = \ \frac{1}{z^2} \dd z^2 + h_{ij} \dd x^i\dd x^j \, , \label{FGmetric}
\end{align}
and in turn
\begin{equation}
	\mexp_{ij} \ = \ \mexp_{ij}^{0}+z^2 \mexp_{ij}^{2}+z^3 \mexp_{ij}^{3} + o(z^3) \, , \label{metricexp}
\end{equation}
where $\mexp^0_{ij} = g_{ij}$ is the metric on the conformal boundary $(M_3,g)$ at $z=0$.
The volume of $\mexp$ may also be expanded as
\begin{align}
\sqrt{\det \mexp} \ = \ \sqrt{\det \mexp^0} \, \Big[ & 1 + \tfrac{z^2}{2} t^{(2)} + \tfrac{z^3}{2} t^{(3)} \Big] + o(z^3) \, ,
\end{align}
where we have denoted $t^{(n)} \equiv \mathrm{Tr} \left[ (\mexp^0)^{-1} \mexp^n \right]$ and indices are always raised with $\mexp^0$. Just as with the metric, we assume that the other bosonic fields have an analytic expansion near the boundary
\begin{align}
	\label{Xexp}
	X \ =& \ 1 + z X_1 + z^2 X_2 + z^3 X_3 + o(z^3)\, ,   \\[5pt]
	\label{jexp}
	\st \ =& \ z \st_1 + z^2 \st_2 + z^3 \st_3 + o(z^3) \, , \\[5pt]
	\label{aIexp} 
	\cA^I \ =& \ A^I + z \ma_1^I + z^2 \ma_2^I + o(z^2) \, , \\[5pt]
	\label{hataIexp}
	\hat{\cA}^I \ =& \ \hat{A}^I + z \hat{\ma}_1^I + z^2 \hat{\ma}_2^I + o(z^2) \, , 
\end{align}
and we use gauge redundancy to remove the components along $\rd z$ in the gauge fields.

Assuming the perturbative expansions for the bosonic fields, we may then substitute them in the corresponding equations of motion \eqref{Xeom}--\eqref{geom} and order by order find relations between the coefficients. The results (up to the relevant order) are summarized here:
\begin{align}
	\g^2 \ =& \ \tfrac{1}{2} \, , \\[5pt]
	\label{LapAxion1}
	\nabla^2 \st_1  \ =& \ \st_1 ( t^{(2)} + 2 X_1^2 + 4 X_2 ) + 4 X_1 \st_2 + 2 \st_3 \, , \\[5pt]
	\label{LapDilaton1}
	\nabla^2 X_1 \ =& \ 2 X_3 + X_1 ( t^{(2)} + 2 X_1^2 - 2 X_2 + \st_1^2 ) + \st_1 \st_2 - 2 \st_1 ( X_1 \st_1 + \st_2 )  \, , \\[5pt]
	\label{aI2eqn}
	0 \ =& \ D *_{\mexp^0} \ma_1^I \, , \qquad \ma_2^I \ = \ X_1 \ma_1^I + \tfrac{1}{2} *_{\mexp^0} D *_{\mexp^0} F^I - \tfrac{\ii}{2} \st_1 *_{\mexp^0} F^I \, , \\[5pt]
	\label{hataI2eqn}
	0 \ =& \ \hat{D} *_{\mexp^0} \hat{\ma}_1^I \, , \qquad \hat{\ma}_2^I \ = \ - X_1 \hat{\ma}_1^I + \tfrac{1}{2} *_{\mexp^0} \hat{D} *_{\mexp^0} \hat{F}^I + \tfrac{\ii}{2} \st_1 *_{\mexp^0} \hat{F}^I  \, , \\[5pt]
	\label{g2}
	\mexp^2_{ij} \ =& \ - \big[ R_{ij}(\mexp^0) - \tfrac{1}{4} \mexp^0_{ij} R(\mexp^0) \big] - \mexp^0_{ij} \big( \tfrac{1}{2} X_1^2 + \tfrac{1}{8} \st_1^2 \big) \, ,  \\[5pt]
	\label{t3}
	t^{(3)} \ =& \ \tfrac{4}{3} X_1^3 - \tfrac{2}{3} X_1 ( 4 X_2 + \st_1^2 ) - \tfrac{2}{3} \st_1 \st_2 \, , \\[5pt]
	\label{eq:divg3}
	\tfrac{3}{2}\nabla^i \mexp^3_{ij} \ =& \ - \j_1^2 \nabla_j X_1 - \tfrac{1}{2}\j_1 \nabla_j \j_2 + 2 X_1^2 \nabla_j X_1 - 2 X_1 \nabla_j X_2 + \tfrac{1}{2}F^I_{ji}({\rm a}^I_1)^i + \tfrac{1}{2}\hat{F}^I_{ji}(\hat{{\rm a}}^I_1)^i
	\, .
\end{align}
The curvatures appearing are 
\beq
F^I \ \equiv \ \dd A^I + \tfrac{1}{2} \gc\, \epsilon^{IJK} A^J \wedge A^K \, , \qquad \hat{F}^I \ \equiv \ \dd \hat{A}^I + \tfrac{1}{2} \gc\, \epsilon^{IJK} \hat{A}^J \wedge \hat{A}^K \, .
\eeq
Taking the trace of \eqref{g2}, we immediately find
\begin{align}
	\label{t2}
	t^{(2)} \ =& \ - \tfrac{1}{4} R(\mexp^0) - \tfrac{3}{2} X_1^2 - \tfrac{3}{8} \st_1^2 \, .
\end{align}

\subsection{Holographic renormalization}
\label{subsec:HoloRen}

Our aim is to compare a gravity observable with its CFT boundary dual, and more specifically we want to compute the dual to the CFT effective action. However, due to the presence of scalar fields in the bulk, the problem may present two subtleties: one related to the renormalization scheme, and the second one related to the AdS/CFT dictionary itself. 

\medskip

First, we begin by renormalizing the on-shell action for an arbitrary asymptotically locally AdS solution $(Y_4,G)$ using holographic renormalization \cite{Emparan:1999pm, Taylor:2000xw, deHaro:2000vlm}. To do so, we cutoff $Y_4$ at a small radial distance $z=\delta$ from the (conformal) boundary, obtaining a space $Y_\delta$ with a boundary $M_\delta \cong M_3$ with induced metric $h$ (compare with \eqref{FGmetric}). As required by the presence of a boundary, we need to add a Gibbons--Hawking--York term in order to obtain the correct equations of motion. Therefore, the value of the on-shell action for the cutoff space is
\beq
\begin{split}
I_{\rm o-s} + I_{\rm GHY} =  \frac{1}{2\kappa_4^2} \int_{Y_\delta} \ \Big[ &- ( 4 + X^2 + \tilde{X}^2 ) \, \vol_G - \tfrac{1}{2} X^{-2} \big( \cF^I \wedge * \cF^I + \ii \st X^2 \cF^I \wedge \cF^I \big) \\
	&- \tfrac{1}{2} \tilde{X}^{-2} \big( \hat{\cF}^I \wedge * \hat{\cF}^I - \ii \st X^2 \hat{\cF}^I \wedge \hat{\cF}^I \big)\Big] - \frac{1}{\kappa_4^2} \int_{M_\delta} K \, \vol_h \, .
\end{split}
\eeq
This quantity diverges as we take the cutoff $\delta \to 0$. In order to cancel the divergences, we can include divergent counterterms constructed out of the induced geometry on $M_\delta$. The simplest such counterterms are \cite{GRS1}
\begin{align}
\label{eq:Rct}
I_{\rm ct,g} \ =& \ \frac{1}{\kappa^2_4}\int_{M_\delta} \frac{1}{2}R(h) \, \vol_h \, , \\
\label{eq:CTscalars}
I_{\rm ct,s} \ =& \ \frac{1}{\kappa^2_4} \int_{M_\delta} \left( 2+(X-1)^2 + \frac{1}{4}\j^2 \right) \, \vol_h \, .
\end{align}
Adding these counterterms guarantees that the limit
\beq
\label{eq:Snaive}
\S \ = \ \lim_{\delta\to 0}\left[ I_{\rm o-s} + I_{\rm GHY} + I_{\rm ct,g} + I_{\rm ct,s} \right] \, ,
\eeq
is finite.

However, it is not possible to uniquely fix the functional form of the scalar counterterm $I_{\rm ct, s}$ using only its divergent behaviour. There are potentially infinitely many counterterms that make \eqref{eq:Snaive} finite, each differing by finite terms, corresponding to different regularization schemes. In our case, we would like to find a regularization scheme that preserves the rigid supersymmetry of the boundary, making the resulting $\mathbb{S}^{\rm SUSY}$ invariant under supersymmetry provided the appropriate transformation of the sources. The correct counterterm that achieves this is
\beq
\label{eq:properCT}
I_{\rm ct, s}^{\rm SUSY} \ = \ \frac{1}{\kappa^2_4}\int_{M_\delta}2\mc{W} \, \vol_h \ = \ \frac{1}{\kappa^2_4}\int_{M_\delta}\sqrt{2+X^2+\tilde{X}^2} \, \vol_h \, ,
\eeq
where $\mc{W}$ is the $\mc{N}=1$ (Lorentzian) superpotential discussed in Appendix \ref{app:Reduction}. The necessity of this counterterm has been argued for in similar models in \cite{FreedmanPufu, FreedmanPufuEtAl, Halmagyi:2017hmw, Cabo-Bizet:2017xdr, Gauntlett:2018vhk, Bobev:2018wbt, Arav:2018njv, Bobev:2020pjk}. We provide arguments for the $\mc{N}=4$ supergravity in Appendices \ref{app:Reduction} and \ref{app:BdrySUSY}.

\medskip

Our discussion gives us an on-shell action holographically renormalized preserving supersymmetry: 
\beq
\label{eq:SSusy}
\S^{\rm SUSY} \ = \ \lim_{\delta\to 0}\left[ I_{\rm o-s} + I_{\rm GHY} + I_{\rm ct,g} + I_{\rm ct,s}^{\rm SUSY} \right] \, .
\eeq
This leaves us with the second subtlety. As we mentioned, both bulk scalars $\phi, \j$ have mass $m^2 = - 2$. The AdS/CFT relation between the mass of a bulk scalar and the scaling dimension of the dual CFT operator \eqref{eq:AdSCFTScaling} then gives the two possibilities $\Delta_+ = 2, \Delta_- = 1$. As long as the bulk scalars are dual to boundary operators with scaling dimension $\Delta_+ = 2$, we could go on with the ``standard'' application of the AdS/CFT dictionary, and identify the gravitational free energy in \eqref{eq:AdSCFT} with $\mathbb{S}^{\rm SUSY}$, so that the we can interpret the gravitational on-shell action as the generator of connected diagrams for the (deformed) CFT as a function of the sources.

Here instead we are working with a boundary SCFT containing a scalar with dimension $\Delta_- = 1$, and this behaviour is generic in three dimensions with $\mc{N}\geq 4$ supersymmetry. In this case, as explained in Section \ref{sec:Intro}, we cannot interpret the on-shell action as the SCFT generating functional, but rather its Legendre transform with respect to the bulk scalar dual to the dimension $1$ operator \cite{Klebanov:1999tb}. 

Concretely, the bulk field corresponding to the dimension $2$ operator in the stress-energy tensor multiplet \eqref{eq:3dTmultiplet} is the pseudoscalar $\j$. Looking at its Fefferman--Graham expansion \eqref{jexp}, we interpret $\j_1$ as a source for the dual operator. The bulk equations of motion are solved with Dirichlet boundary conditions for $\j$, and the on-shell action is interpreted as a functional of $\j_1$.

The bulk field corresponding to the dimension $1$ operator, which has to be the scalar $X$, should instead obey alternate boundary conditions. Already from its expansion \eqref{Xexp}, we see that on the boundary we cannot construct a scale-invariant deformation of the usual form
\beq
\int_{M_3} X_1 \Xop \, \vol_g \, .
\eeq
Instead, we should view the source for $\Xop$ as the variable canonically conjugate to the leading coefficient $X_1$
\beq
\Xsource \ = \ \frac{1}{\sqrt{g}}\frac{\delta \S^{\rm SUSY}}{\delta X_1} \ = \ \frac{1}{\kappa^2_4}\left( X_1^2 -2 X_2 + \tfrac{1}{2}\j_1^2 \right) \, ,
\eeq
which is the sub-leading term in the expansion with non-linear corrections due to supersymmetry. We then perform a Legendre transformation on $\S^{\rm SUSY}$
\beq
\label{eq:LegendreTransformation}
\tilde{\S} \ = \ \S^{\rm SUSY} - \int_{M_3} \Xsource X_1 \, \vol_g \, ,
\eeq
viewed as a functional of $\Xsource$ instead of $X_1$, and obtained after extremization of the right-hand side with respect to $X_1$. 

It is $\tilde{\S}$ that should be interpreted as the generating functional for the connected diagrams in the dual SCFT: a functional of the sources represented by the boundary conditions for the bulk fields $(g_{ij}, \Xsource, \j_1, A_i^I, \hat{A}_i^I)$. We may then compute the VEVs for the boundary dual operators holographically
\begin{align}
	\label{Tij}
	\langle T_{ij} \rangle \ =& \ \frac{2}{\sqrt{g}}\frac{\delta \tilde{\S}}{\delta g^{ij}} \ = \ \frac{1}{\kappa_4^2} \Big[ \tfrac{3}{2} \mexp^3_{ij} - \tfrac{1}{2} g_{ij} \big( 3 t^{(3)} + 8 X_1 X_2 - 4X_1^3 + \st_1 \st_2 \big) \Big]  \, , \\[5pt]
	\label{Xi}
	\langle \Xop \rangle \ =& \ \frac{1}{\sqrt{g}}\frac{\delta \tilde{\S}}{\delta \Xsource} \ = \ - X_1 \, , \\[5pt]
	\label{Sigma}
	\langle \jop \rangle \ =& \ \frac{1}{\sqrt{g}}\frac{\delta \tilde{\S}}{\delta \j_1} \ = \ - \frac{1}{\kappa_4^2} \big( X_1 \st_1 + \tfrac{1}{2} \st_2 \big) \, , \\[5pt]	
	\label{calJi}
	\langle \mathscr{J}^{I}_i \rangle \ =& \ \frac{1}{\sqrt{g}}\frac{\delta \tilde{\S}}{\delta A^I_i} \ = \ -\frac{1}{2\kappa_4^2}(\ma_1^I)_i~,\\[5pt]
	\label{bbJ}
	\langle \hat{\mathscr{J}}^{I}_i \rangle \ =& \ \frac{1}{\sqrt{g}}\frac{\delta \tilde{\S}}{\delta \hat{A}^{ij}} \ = \ -\frac{1}{2\kappa_4^2}(\hat{\ma}^I_1)_i \, .
\end{align}
Note that each of these expressions contains terms which are not expressible as boundary quantities via the Fefferman--Graham expansion of the previous section. Nevertheless, relations between them can be found by deriving boundary Ward identities from the bulk. This will provide some consistency checks.

\medskip

Under a variation of the boundary data, $\tilde{\S}$ varies as
\beq
\label{eq:Variation}
\delta \tilde{\S} \ = \ \int_{M_3} \left[ \tfrac{1}{2} \langle T_{ij} \rangle \delta g^{ij} + \langle \Xop \rangle \delta \Xsource + \langle \jop \rangle \delta \j_1 + \langle \mathscr{J}^I_i \rangle \delta A^{Ii} + \langle \hat{\mathscr{J}}^I_i \rangle \delta \hat{A}^{Ii} \right] \vol_g \, .
\eeq
Choosing a boundary Weyl rescaling, under which
\beq
\begin{aligned}
\delta g^{ij} \ &= \ -2\sigma g^{ij} \, , \qquad & \delta A^{Ii} \ &= \ \delta \hat{A}^{Ii} \ = \ 0 \, , \\
\delta \Xsource \ &= \ - (3-\Delta_-) \sigma \Xsource \, , \quad & \delta \j_1 \ &= \ - (3-\Delta_+) \sigma \j_1
 \, ,
\end{aligned}
\eeq
compatibly with \eqref{eq:3dTmultiplet}, we find that the expressions \eqref{Tij}--\eqref{Sigma} satisfy the following relation
\beq
\langle T^i_{\ph{i}i} \rangle \ = \ -2 \langle \mc{O}_{\Delta_-} \rangle \Xsource - \langle \mc{O}_{\Delta_+} \rangle \j_1 \, ,
\eeq
corresponding to the vanishing of the Weyl anomaly in three dimensions and the standard deformation of the trace of the stress-energy tensor by operators in a CFT.
This confirms the consistency of the scaling of the operators and the necessity of the Legendre transform; the other choice of quantization and scalings (namely both scalar and pseudoscalar having scaling dimension $2$) gives a vanishing conformal anomaly when derived using just $\mathbb{S}$ (without supersymmetry-preserving counterterms and Legendre transformation) \cite{GRS1}.

A boundary gauge transformation of the form $\delta A^I_i = D_i\theta^I$ gives the conservation equations for the boundary $R$-symmetry currents
\beq
0 \ = \ *D * \mathscr{J}^I \ = \ *D* \hat{\mathscr{J}}^I \, ,
\eeq
which are equivalent to the constraints \eqref{aI2eqn} and \eqref{hataI2eqn}. Similarly, invariance of $\tilde{\mathbb{S}}$ under a generic boundary diffeomorphism generated by the vector field $\xi^i$ gives the conservation equation for the stress-energy tensor
\beq
\nabla^j \langle T_{ji} \rangle \ = \ - \langle \mc{O}_{\Delta_-} \rangle \nabla_i \Xsource - \langle \mc{O}_{\Delta_+} \rangle \nabla_i \j_1 - F^I_{ij} \langle \mathscr{J}^{Ij} \rangle - \hat{F}^I_{ij} \langle \hat{\mathscr{J}}^{Ij} \rangle \, ,
\eeq
which is equivalent to the equation \eqref{eq:divg3} obtained from the expansion of the $zi$ component of the Einstein equation of motion \eqref{geom}. Perhaps less standard is the Ward identity corresponding to a boundary supersymmetry variation of $\tilde{\mathbb{S}}$: we consider it in Appendix \ref{app:SUSYWard}. 

\section{The twist in the bulk}
\label{sec:BulkTwist}

In this section, we examine of the consequences of the topological twist in the bulk. We begin by applying the Fefferman--Graham expansion to the supersymmetry equations, showing that at the boundary we indeed recover the equations of off-shell $3d$ conformal supergravity. With the knowledge obtained from the Fefferman--Graham expansion, we then show that for any supersymmetric asymptotically locally AdS solution, the functional $\tilde{\mathbb{S}}$ is independent of the boundary metric provided we apply the boundary conditions of the topological twist. Finally, we compute the value of the topological invariant for smooth solutions, showing once more the importance of the supersymmetric counterterms and of the Legendre transformation.

\subsection{Expansion of the supersymmetry equations}
\label{subsec:ExpansionSUSY}

The same Fefferman--Graham expansion applied in the previous section to the bosonic fields may also be done for the Killing spinor
\beq
\epsilon^a \ = \ z^{-1/2} \, \varepsilon^a + z^{1/2} \, \xi^a + o(z^{1/2}) \, .
\eeq
The expansion of the gravitino equation \eqref{eq:BulkGravitino} along $z$ direction gives, at the two leading orders
\beq
0 = z^{-1/2} \left( \Gamma_{\overline{z}} + \sqrt{2}\g \identity \right) \varepsilon^a + z^{1/2} \left[ \tfrac{1}{2} \left( \identity - \sqrt{2}\g \Gamma_{\overline{z}} \right) \xi^a + \tfrac{\ii}{4} \j_1 \Gamma_5 \left( \identity + \sqrt{2}\g \Gamma_{\overline{z}} \right) \varepsilon^a \right] + o(z^{1/2}) \, .
\eeq
We can solve these constraints by imposing that $\varepsilon^a$ and $\xi^a$ have definite chirality under $\Gamma_{\overline{z}}$. Appealing to the $\Z_2$ symmetry of the equations of motion and supersymmetry variations that sends $\g \to - \g$, $(A^I_\mu, \hat{A}^I_\mu) \to (- A^I_\mu, -\hat{A}^I_\mu)$, $\Gamma_\mu \to - \Gamma_\mu$, we can choose $\g = - \tfrac{1}{\sqrt{2}}$ and then $\varepsilon^a$ ($\xi^a$) has positive (negative) chirality under $\Gamma_{\overline{z}}$.

In order to describe the boundary structure of the supersymmetry, we need to decompose four-dimensional spinors and Clifford algebra into their three-dimensional counterparts. To do so, we introduce the basis
\beq
\Gamma_{\overline{1}} \ = \ \Gamma_{\overline{z}} \ = \ \begin{pmatrix}
\identity_2 & 0 \\ 0 & - \identity_2
\end{pmatrix} \, , \qquad \Gamma_{\overline{i+1}} \ = \ \begin{pmatrix}
0 & \sigma_i \\ \sigma_i & 0
\end{pmatrix} \, , \qquad 
\Gamma_5 \ = \ \begin{pmatrix}
0 & - \ii \identity_2 \\ \ii \identity_2 & 0
\end{pmatrix} \, ,
\eeq
and we can write
\beq
\epsilon^a \ = \ z^{-1/2} \begin{pmatrix}
\varepsilon^a_L \\ 0
\end{pmatrix} + z^{1/2} \begin{pmatrix}
0 \\ \xi^a_R
\end{pmatrix} + o(z^{1/2}) \, ,
\eeq
where $L$ and $R$ indicate the chirality with respect to $\Gamma_{\overline{z}}$

Using these results, the gravitino equation \eqref{eq:BulkGravitino} along the directions tangent to $M_3$ becomes
\beq
\label{eq:LeadingOrderGravitino}
0 \ = \ \nabla_i \varepsilon^a_L + \tfrac{1}{2\sqrt{2}} \, \eta^I_{ab}A^I_i \, \varepsilon^b_L - \tfrac{1}{2\sqrt{2}} \, \overline{\eta}^I_{ab}\hat{A}^I_i \, \varepsilon^b_L - \tfrac{1}{4} \j_1 \, \sigma_i \varepsilon^a_L + \sigma_i \xi^a_R \, ,
\eeq
whereas the dilatino equation \eqref{eq:BulkDilatino} gives two different equations on the two subspaces corresponding to the two chiralities of $\Gamma_{\overline{z}}$:
\begin{align}
\label{eq:LeadingOrderDilatino1}
\begin{split}
0 \ &= \ -\tfrac{1}{2\sqrt{2}}\left( X_1^2 - 2X_2 \right) \varepsilon^a_L - \tfrac{1}{\sqrt{2}}\j_1 \, \xi^a_R + \tfrac{1}{2\sqrt{2}}\partial_i \j_1 \, \sigma^i\varepsilon^a_L  + \tfrac{1}{8} \left( \eta^I_{ab}F^I_{ij} + \overline{\eta}^I_{ab}\hat{F}^I_{ij} \right) \sigma^{ij}\varepsilon^b_L  \, ,
\end{split} \\[5pt]
\label{eq:LeadingOrderDilatino2}
\begin{split}
0 \ &= \ - \tfrac{1}{2\sqrt{2}}\left( \j_2 + X_1\j_1\right) \varepsilon^a_L - \sqrt{2} X_1 \xi^a_R + \tfrac{1}{\sqrt{2}}\partial_i X_1 \sigma^i \varepsilon^a_L - \tfrac{1}{4}\left( \eta^I_{ab}(\ma_1^I)_i + \overline{\eta}^I_{ab}(\hat{\ma}^I_1)_i \right) \sigma^i \varepsilon^b_L \, .
\end{split}
\end{align}

It's clear that \eqref{eq:LeadingOrderGravitino} reproduces the gravitino variation of the three-dimensional off-shell conformal supergravity \eqref{eq:3dSUGRAGravitino} provided we identify
\beq
\label{eq:Identification1}
(\psi^a_i)^{3d} \ = \ \psi^a_{0iL} \, , \qquad \qquad \varepsilon^a_L \ = \ \tredSUGRAQ^a \, , \qquad \qquad \xi^a_R \ = \ \tredSUGRAS^a + \tfrac{1}{4} \j_1 \tredSUGRAQ^a \, .
\eeq
When we substitute this into the equations coming from the dilatino, and use the relations obtained from holographic renormalization \eqref{Tij}--\eqref{bbJ}, we have
\begin{align}
\label{eq:BdryDilatino1}
\begin{split}
0 \ &= \ \partial_i \j_1 \, \sigma^i\tredSUGRAQ^a - \kappa^2_4 \Xsource \, \tredSUGRAQ^a + \tfrac{1}{2\sqrt{2}} \left( \eta^I_{ab}F^I_{ij} + \overline{\eta}^I_{ab}\hat{F}^I_{ij} \right) \sigma^{ij}\tredSUGRAQ^b + \j_1 \tredSUGRAS^a \, ,
\end{split} \\
\label{eq:BdryDilatino2}
\begin{split}
\delta \chi^a_{0R} \ &= \ \tfrac{1}{\sqrt{2}} \Big[ \kappa^2_4 \langle \mc{O}_{\Delta_+} \rangle \tredSUGRAQ^a - \partial_i \langle \mc{O}_{\Delta_-} \rangle \, \sigma^i \tredSUGRAQ^a  + 2 \langle \mc{O}_{\Delta_-} \rangle \tredSUGRAS^a \\
&\ \qquad \quad + \tfrac{\kappa^2_4}{\sqrt{2}} \left( \langle \mathscr{J}^I_i \rangle \eta^I_{ab} + \langle \hat{\mathscr{J}}^I_i \rangle \overline{\eta}^I_{ab} \right) \sigma^i \tredSUGRAQ^b \Big] \, .
\end{split}
\end{align}
Again, we see that \eqref{eq:BdryDilatino1} corresponds to \eqref{eq:3dSUGRAConstraint} upon identifying
\beq
\label{eq:Identification2}
(\chi^a)^{3d} \ = \ \sqrt{2} \chi^a_{0L} \, , \qquad  \Xsource \ = \ - \frac{1}{\kappa^2_4} S_2 \, , \qquad \j_1 \ = \ 2S_1 \, .
\eeq
which is consistent with the fact that $\j_1$ should be interpreted as the source for the operator $\mc{O}_{\Delta_+}$ and $\Xsource$ as the source for the operator $\mc{O}_{\Delta_-}$. The second equation \eqref{eq:BdryDilatino2} has the same structure as the first one, but with sources and VEVs exchanged (notice that $\sigma^{ij}$ can be dualized together with the $R$-symmetry currents, and $*\mathscr{J}^I$ is closed, as a curvature should be). It represents a BPS condition among the VEVs of the operators of the SCFT coupled to the $3d$ off-shell supergravity. As such, we can reproduce it by considering the holographic Ward identity corresponding to the supersymmetry of the boundary theory, as discussed in Appendix \ref{app:SUSYWard}. Thus, at the boundary of an asymptotically locally AdS solution of $Spin(4)$ supergravity we find off-shell conformal supergravity, and we will present further evidence in Appendix \ref{app:BdrySUSY}. This is consistent with the usual lore, and provides additional support to the expectations that had been used in \cite{Nishimura:2012jh} in order to construct $\mc{N}=8$ off-shell conformal supergravity.

\medskip

We may now solve \eqref{eq:LeadingOrderGravitino} and \eqref{eq:BdryDilatino1} on an arbitrary three-manifold by setting the topological twist conditions \eqref{eq:TopTwistConnection} and \eqref{eq:TopTwistSUGRA}, which we rephrase now in terms of the gravity fields
\beq
\label{eq:BCTwist}
A^I_i \ = \ \frac{1}{\sqrt{2}}\epsilon^I_{\ph{I}\overline{jk}}\omega_i^{\ph{i}\overline{jk}} \, , \qquad \j_1 \ = \ 0 \, , \qquad \Xsource \ = \ \frac{1}{2\kappa^2_4}R \, ,
\eeq
where $\omega_i^{\ph{i}\overline{jk}}$ and $R$ refer to the boundary metric $g$, and we consistently truncate the theory to the sector with $\hat{\mc{A}}^I_i \equiv 0$. Furthermore, the additional constrain \eqref{eq:BdryDilatino2} provides the following non-trivial relation between the sub-leading terms in the expansions of the bosonic fields (or, equivalently, between the VEVs of the dual operators)
\beq
\label{eq:BCTwist2}
\begin{split}
\delta^{\overline{i}}_I \langle \mathscr{J}^I_{\overline{i}} \rangle  \ =  - \frac{\sqrt{2} \ii}{\kappa^2_4}\langle \mc{O}_{\Delta_+} \rangle \,  \qquad &\Leftrightarrow \qquad \ \ \delta^{\overline{i}}_I\left( \ma^I_1 \right)_{\overline{i}} \ = \ - \ii \sqrt{2}\, \j_2 \, , \\
\epsilon_{\overline{ij}I}\langle \mathscr{J}^{I\overline{j}} \rangle \ = \ \frac{\sqrt{2}}{\kappa^2_4} \partial_{\overline{i}} \langle \mc{O}_{\Delta_-} \rangle \qquad & \Leftrightarrow \qquad \epsilon_{\overline{ij}I}\left(\ma^I_1\right)^{\overline{j}} \ = \ 2\sqrt{2} \, \partial_{\overline{i}}X_1 \, .
\end{split}
\eeq
At first sight it may seem that these assignments violate reality conditions. However, as discussed in \cite{GRS1}, in the truncated sector of the theory where $\hat{\mc{A}}^I \equiv 0$, it is consistent to impose that the bulk Killing spinor $\epsilon^a$ is symplectic Majorana, and all the fields are real except for $\j$ which is imaginary.

\medskip

Having set the boundary conditions using the topological twist, we continue to expand the bulk gravitino and dilatino equations to higher orders in order to obtain further conditions imposed by supersymmetry on the fields not fixed by the boundary data. The result relevant to our purposes is the expression for the subleading metric
\beq
	\label{g3}
	\mexp^3_{\overline{ij}} \ = \ \tfrac{2}{3} \nabla_{\overline{i}}\nabla_{\overline{j}}X_1 + \tfrac{2}{3} X_1  R_{\overline{ij}} + \tfrac{1}{6\sqrt{2}} (F_{1(\overline{i}})^{\overline{kl}}\epsilon_{\overline{j})\overline{kl}} - \tfrac{1}{3\sqrt{2}}(F_1^{\overline{k}})^{\overline{l}}_{\ph{l}( \overline{i}}\epsilon_{\overline{j})\overline{kl}} \, ,
\eeq
where
\beq
\label{eq:FI1}
F^I_1 \ = \ D\ma^I_1 \ \equiv \ \rd \ma^I_1 - \frac{1}{\sqrt{2}}\epsilon^{IJK}A^J\wedge \ma^K_1 \, .
\eeq

\subsection{Variation of the action}
\label{subsec:Variation}
 
In this short section we consider the (Legendre transform of the) action of asymptotically locally AdS solutions with boundary conditions given by the topological twist, and we show that it does not depend on the boundary metric. We start from the variation \eqref{eq:Variation}, but now we observe that the twist boundary conditions \eqref{eq:BCTwist}, \eqref{eq:BCTwist2} mean that the variations of the boundary data $(g_{ij}, \Xsource, \j_1, A^I_i)$ can all be related to $\delta g^{ij}$, since all the boundary data are fixed in terms of $g_{ij}$.

Using standard formulae from Riemannian geometry (reviewed in \cite{GRS1}), and dropping total derivatives which vanish for the closed manifolds $M_3$ that we consider, we find that we can write
\beq
\label{eq:Variationg}
\delta \tilde{\mathbb{S}} \ = \ \frac{1}{2\kappa^2_4}\int_{M_3} \frac{1}{2}\mc{T}_{ij} \delta g^{ij} \, \vol_g
\eeq
with the effective stress-energy tensor
\beq
\mc{T}_{ij} \ = \ 3 \mexp^3_{ij} + 2 \left( - X_1 R_{ij} + \nabla_i\nabla_j X_1 - \nabla^2 X_1 \, g_{ij} \right) + \sqrt{2} \nabla^k \left( \epsilon^I_{\ph{I}k(i}(a^I_1)_{j)} \right) \, .
\eeq
The (Legendre transform of the) on-shell gravitational action is invariant under changes in the boundary metric, as we expect from field theory, if the effective stress-energy tensor vanishes, $\mc{T}_{ij} \equiv 0$ for any Riemannian three-manifold.

Substituting the expression for $\mexp^3_{ij}$ found from the expansion of the supersymmetry variations, we have
\beq
\label{eq:Teff_v1}
\begin{split}
\mc{T}_{ij} &= 4 \nabla_{i}\nabla_{j}X_1 - 2\nabla^2 X_1 \, g_{ij} + \tfrac{1}{2\sqrt{2}} (F_{1({i}})^{{kl}}\epsilon_{{j}){kl}} - \tfrac{1}{\sqrt{2}}(F_1^{{k}})^{{l}}_{\ph{l}( {i}}\epsilon_{{j}){kl}} + \sqrt{2} \nabla^k \left( \epsilon^I_{\ph{I}k(i}(a^I_1)_{j)} \right) \, ,
\end{split}
\eeq
and we already observe that the terms linear in $X_1$ have simplified. In order to continue, it is useful to observe the following fact. As we have already stressed, to perform the topological twist we identify the $R$-symmetry bundle with the tangent bundle of the three-manifold, further equating their connections as in \eqref{eq:BCTwist}. This means that $(\ma^I_1)_i$, rather than being a one-form (connection) valued in the adjoint of the gauge algebra should really be viewed as a $(1,1)$ tensor on $M_3$. Therefore, the covariant derivative in \eqref{eq:Teff_v1} acts on all the indices of the tensor, and we should also review the definition of $(F_1^I)_{ij}$: because of the identification of the connections \eqref{eq:BCTwist}, from its definition \eqref{eq:FI1}, we see that it really is the antisymmetrization of two indices of the full covariant derivative of $(\ma^I_1)_i$
\beq
 F^{\ph{1}I}_{1\ph{I}ij} \ = \ 2 \nabla_{[i} \ma_{1\ph{I}j]}^{\ph{1}I} \, .
\eeq
In fact, more is true because the topological twist fixes the antisymmetric part of $\ma_1$ as written in \eqref{eq:BCTwist2}, so
\beq
F^{\ph{1}I}_{1\ph{I}ij} \ = \ 2 \nabla_{[i}\ma_{1(j] I)} - 2\sqrt{2} \epsilon_{Ik[i}\nabla_{j]}\nabla^k X_1 \, .
\eeq
Thus, \eqref{eq:Teff_v1} ends up containing only terms with covariant derivatives of the symmetric part of $\ma_1$ and double derivatives acting on $X_1$. After a careful expansion, we ultimately find that $\mc{T}_{ij}=0$. Notice that this holds for any boundary data $(M_3, g)$, independently of the precise expression of $\ma_1$ and $X_1$, which are only fixed by the regularity of the bulk solution and not by the Fefferman--Graham expansion.

\medskip

It is important to spell out some of the subtler assumptions in our derivation. This is based on the validity of the expression \eqref{eq:Variation} for the variation $\delta \tilde{\mathbb{S}}$ (which in turn determines the effective stress-energy tensor in \eqref{eq:Variationg}). Generically, varying the boundary data results in a change of the bulk fields. However, we are evaluating $\tilde{\mathbb{S}}$ on a solution to the equations of motion, which means that by definition the bulk variation vanishes and the only contribution can come from the boundary $(M_3,g)$. Should there be any additional internal boundaries or singularities in the solution, these would contribute to the variation \eqref{eq:Variationg}. If one does not require smoothness of the solution, then we can interpret the computation in this section as providing constraints on the allowed singularities and boundary conditions in the interior: they should not contribute to \eqref{eq:Variationg}, or else $\tilde{\mathbb{S}}$ would not be a topological invariant.

\medskip

We should also emphasise the importance of the choice of supersymmetry-preserving renormalization scheme. The boundary supersymmetry guarantees that the field theory partition function only depends on a subset of the geometric data specifying the rigid supersymmetric background; in this case it should be independent of the metric. However, this statement relies on the absence of ``supersymmetry anomalies.'' In the four-dimensional case, this problem is solved from the start, because Witten's topologically twisted $\mc{N}=2$ theory reproduces the Donaldson invariants of the background manifold \cite{Witten:1988ze}, which are proved to only depend on its diffeomorphism class. This is paralleled in the dual gravity computation by the fact that the invariance of the on-shell action follows in the minimal holographic renormalization scheme \cite{BenettiGenolini:2017zmu}. However, the minimal holographic renormalization scheme is not supersymmetry-preserving for the duals to four-dimensional $\mc{N}=1$ theories, as shown by a computation analogous to the one presented here \cite{Genolini:2016sxe, Genolini:2016ecx}.\footnote{The field-theoretic statement, based on supersymmetry, is that the partition function depends holomorphically on the complex structure of the underlying complex four-manifold \cite{Closset:2013vra}. In fact, this statement is scheme dependent, as it depends on the absence of 't Hooft anomalies for the flavor symmetries \cite{Closset:2019ucb}. This also shows that ``supersymmetry anomalies'' in the field theory formulated on a curved background first pointed out in \cite{Papadimitriou:2017kzw, An:2017ihs} can be removed by local counterterms \cite{Closset:2019ucb, Kuzenko:2019vvi}.} 

In the three-dimensional case considered here, the dual supergravity theory has scalars that can be quantized in two ways. Choosing the dual operators to both have scaling dimension $2$ is not consistent with the boundary supersymmetry, which instead dictates the choice of alternate quantization for the bulk scalars, and thus requires the bulk observables to be regulated in a way that is consistent with supersymmetry, forcing us to add the counterterm \eqref{eq:properCT}. This guarantees that the gravitational free energy, the Legendre transform of the renormalized on-shell action, is independent of the boundary metric.

\subsection{On-shell action}
\label{subsec:OSAction}

Finally, we compute the value of $\tilde{\mathbb{S}}$ with twisted boundary conditions for any smooth supergravity solution. We will be succinct, and we refer the reader to \cite{GRS1} for additional details.

\medskip

In the truncated sector of the supergravity theory where $\hat{\mc{A}}^I_i \equiv 0$ (which is the one relevant for the twist), it is consistent to impose that the bulk Killing spinor $\epsilon^a$ is symplectic Majorana, and further project onto a space of definite chirality with respect to the action of $(\Gamma_5)^a_{\ph{a}b}$ on the $R$-symmetry indices. Effectively, each of the four components of the bulk spinor may be related to a single Dirac spinor $\zeta$.

A single Dirac spinor in four dimensions defines a local identity structure, that is, a local orthonormal frame $\{ \E^1, \dots, \E^4\}$ and two scalar functions $S, \theta$, constructed out of spinor bilinears.
The frame $\{ \E^1, \dots, \E^4\}$ degenerates where the spinor vanishes and where it becomes chiral, so we define $Y^{(0)}_4$ to be the subset of $Y_4$ where this does not happen. We should then ask ourselves about the global definition of the frame on $Y^{(0)}_4$.

Globally, $\epsilon^a$ is (formally) a section of $Spin(Y_4)\otimes E$, where $E$ is a real rank-$4$ vector bundle associated to the principal $R$-symmetry $SU(2)_R$ bundle. 
The most generic background admitting a globally well-defined bulk Killing spinor $\epsilon^a$ is $Y_4^{(0)}$ with a $Spin^{SU(2)}(4)$ structure, where\footnote{These structures were originally defined in \cite{Back:1978zf, Avis:1979de}.} 
\beq
Spin^{SU(2)}(4) \ \equiv \ \frac{Spin(4) \times SU(2)}{\Z_2} \, .
\eeq

\medskip

From the bulk Killing spinor equations \eqref{eq:BulkGravitino} and \eqref{eq:BulkDilatino} follows a set of spinor equations for $\zeta$, and from those we can use standard spinor bilinears to find a set of differential equations for the frame $\{ \E^1, \dots, \E^4\}$ and the functions $X,S,\j, \theta$. Importantly, having a differential system of global quantities simplifies the computation of the on-shell action. In particular, studying the system of equations implies that in the sector with $\hat{\mc{A}}^I \equiv 0$, the on-shell action is exact
\beq\label{eq:ExactOnShell}
I_{{\rm o-s}} \ = \ - \frac{1}{\kappa^2_4}\int_{Y_4}\dd \left( -\sin\theta \, X^{-1}*\E^4 + X^{-1}*\dd X - \tfrac{1}{2}\j X^4* \dd\j \right)  \, .
\eeq
Thanks to the previous arguments, the three-form whose exterior derivative is being integrated is globally well-defined on $Y_4^{(0)}$. In order to evaluate $I_{{\rm o-s}}$, we first cutoff $Y_4$ near the boundary to $Y_\delta$ (see Section \ref{subsec:HoloRen}), and then surround the loci $Y_4\setminus Y_4^{(0)}$ where the frame degenerates with tubular neighbourhoods of radius $\epsilon$.\footnote{We assume that these loci have measure zero in $Y_4$.} Therefore, we write $I_{\rm o-s}$ as an integral on a space with boundaries of the exterior derivative of a well-defined three-form, and we can apply Stokes' theorem, finding that all the contributions come from the boundaries.

The contributions from the loci where the frame degenerates vanish for a smooth solution as we take the radii of the surrounding tubular neighbourhoods to zero. The analysis necessary to reach this conclusion was carried out in \cite{GRS1}, but concretely only requires us to study the behaviour of the first term in \eqref{eq:ExactOnShell}.

More subtle in this case is the contribution from the UV, where the conformal boundary is. The on-shell action together with the Gibbons--Hawking--York term near the UV reads
\beq
I^{\rm UV}_{\rm o-s} + I_{\rm GHY} \ = \ \frac{1}{\kappa^2_4}\int_{M_3} \left[ - \frac{2}{\delta^2} + \frac{1}{\delta} \left( - \frac{1}{4}R + \frac{1}{2}X_1^2 \right) + \nabla^2 X_1 + \frac{1}{3}R X_1 + o(1) \right] \vol_g \, .
\eeq
Now we may see the importance of the supersymmetric renormalization scheme \eqref{eq:properCT} and of the Legendre transformation \eqref{eq:LegendreTransformation}. Adding the naive minimal counterterms \eqref{eq:Rct} and \eqref{eq:CTscalars} gives
\beq
I_{\rm ct, g} + I_{\rm ct,s} \ = \ \frac{1}{\kappa^2_4}\int_{M_3} \left[ \frac{2}{\delta^3} + \frac{1}{\delta}\left( \frac{1}{4}R - \frac{1}{2}X_1^2 \right) +X_1 \left( X_1^2 + \frac{1}{6}R \right) + o(1) \right] \vol_g \, .
\eeq
Therefore, ignoring total derivatives, we would conclude that
\beq
\mathbb{S} \ = \ \frac{1}{\kappa^2}\int_{M_3} X_1 \left( X_1^2 + \frac{1}{2}R \right) \, \vol_g \, ,
\eeq
which, for generic $X_1$, is not a topological invariant! On the other hand, using the supersymmetry-preserving counterterms \eqref{eq:Rct} and \eqref{eq:properCT} leads to
\beq
I_{\rm ct, g} + I_{\rm ct,s}^{\rm SUSY} \ = \ \frac{1}{\kappa^2_4}\int_{M_3} \left[ \frac{2}{\delta^3} + \frac{1}{\delta}\left( \frac{1}{4}R - \frac{1}{2}X_1^2 \right) +  \frac{1}{6}RX_1 + o(1) \right] \vol_g \, ,
\eeq
and
\beq
\mathbb{S}^{\rm SUSY} \ = \ \frac{1}{\kappa^2}\int_{M_3} \frac{1}{2}RX_1 \, \vol_g \, .
\eeq
Again, this is not a topological invariant, showing very concretely that it cannot be the correct gravitational quantity to be compared with the dual CFT generating functional. On the other hand, inserting the boundary condition \eqref{eq:BCTwist} in the Legendre transformation \eqref{eq:LegendreTransformation} immediately gives that
\beq
\tilde{\mathbb{S}} \ = \ 0 \, ,
\eeq
which is indeed a topological invariant, as we expected. Without any further assumptions on the boundary manifold $M_3$, we conclude that the (Legendre transform of the) on-shell action of any smooth supergravity solution vanishes provided it is renormalized in a way that preserves global supersymmetry.

\section*{Acknowledgments}

We thank James Sparks for helpful discussions and comments on a draft. We have also benefited from discussions with Nikolay Bobev.  
The work of PBG has been partially supported by the Simons Foundation, and by the STFC consolidated grant ST/P000681/1, ST/T000694/1. PR is funded through the STFC grant ST/L000326/1.

\appendix

\section{Reduction from \texorpdfstring{$\mc{N}=8$}{N=8} supergravity}
\label{app:Reduction}

In this appendix we show how to obtain the $\mc{N}=4$ $Spin(4)$ gauged supergravity discussed in the bulk of the paper \cite{DFR} from the maximal $\mc{N}=8$ $Spin(8)$ gauged supergravity \cite{deWit:1982bul}, by extending the ansatz of \cite{Trunc, Gauntlett:2018vhk} to include fermions and gauge fields. This guarantees that we can appeal to the results of \cite{FreedmanPufuEtAl} for the supersymmetry-preserving counterterm, and provides us with the full set of supersymmetry variations, which will be used in the following appendices to study the global supersymmetry of the gravitational free energy. For these purposes it is convenient to work in Lorentzian signature with anticommuting spinors.

\medskip

In four dimensions, the maximal gauged supergravity is $\cN=8$ $Spin(8)$ gauged supergravity. The bosonic fields of this theory are the the metric $G_{\mu\nu}$, $28$ gauge fields $A^{IJ}_\mu$ in the adjoint (antisymmetric) representation, and $35$ complex scalars. It is convenient to represent the scalars by introducing the $56$-bein
\begin{equation}
\label{eq:sechsundfunfzigbein}
\mc V \ \equiv \ 
\begin{pmatrix}
u_{ij}^{~~IJ} & v_{ijKL}
\\
v^{klIJ} & u^{kl}_{~~KL}
\end{pmatrix} \, ,
\end{equation}
where both sets of indices $i,j$ and $I,J$ run from $1$ to $8$, and each pair is antisymmetric. Raising and lowering indices corresponds to taking the complex conjugate. In the fermionic sector, there are eight gravitini $(\psi^i_\mu)^{\cN=8}$ and $56$ dilatini $(\chi^{ijk})^{\cN=8}$ in the relevant spinor representations. Whilst in \cite{deWit:1982bul} the spinors are Weyl spinors and the chirality is linked with their representation of the gauge group, we use Majorana spinors in order to match to the bulk of the paper. To be concrete, the $\Gamma$ matrices $\Gamma^\mu$ generate Cliff$(1,3)$, $\Gamma_5 \equiv \ii \Gamma_{0123}$, and the Majorana conjugate is defined by $\overline{\lambda} \equiv \lambda^T \mathscr{C}$ where $\mathscr{C}$ is the charge conjugation matrix satisfying
\beq
\Gamma_\mu^T \ = \ - \mathscr{C}\Gamma_\mu \mathscr{C}^{-1} \, .
\eeq
A spinor $\lambda$ is Majorana if $\lambda = \lambda^C \equiv \ii \Gamma^0 \mathscr{C}^{-1}\lambda^*$.

\medskip

The reduction consists in looking for a sector invariant under $Spin(4)\times Spin(4)$, and then further truncating. Concretely, we split both lowercase and capital indices $i,I$ into two sets $a,b = 1, \dots, 4$ and $\overline{a}, \overline{b} = 1, \dots 4$ corresponding to two $Spin(4)$ subgroups. Then, we write the following ansatz for the non-vanishing fields
\begin{align}
\label{eq:TruncationGauge}
A^{ab}_\mu \ =& \ \tfrac{1}{2} \eta^I_{ab}\cA^I_\mu - \tfrac{1}{2} \overline{\eta}^I_{ab}\hat{\cA}^I_\mu \, , \\[7pt]
\label{eq:TruncationScalaru}
\begin{split}
u_{ab}^{\ph{ab}cd} \ =& \ \tfrac{1}{2} \left(X + X^{-1}+\ii\varphi X\right)\e^{\ii\theta} \, \delta_{ab}^{cd} \, , \\ 
u_{\overline{ab}}^{\ph{ab}\overline{cd}} \ =& \ \tfrac{1}{2} \left(X + X^{-1}+\ii\varphi X\right)\e^{\ii\theta} \, \delta_{\overline{ab}}^{\overline{cd}} \, , \\
u_{a\overline{b}}^{\ph{ab}c\overline{d}} \ =& \ \tfrac{1}{2} \delta_a^c\delta_{\overline{b}}^{\overline{d}} \, , 
\end{split}\\[7pt]
\label{eq:TruncationScalarv}
\begin{split}
v_{abcd} \ =& \ \tfrac{1}{4} \left( X-X^{-1}-\ii\varphi X \right) \e^{\ii\theta} \, \epsilon_{abcd} \, , \\ 
v_{\overline{abcd}} \ =& \ \tfrac{1}{4} \left( X-X^{-1} + \ii\varphi X \right) \e^{-\ii\theta} \, \epsilon_{\overline{abcd}} \, ,
\end{split}\\[7pt]
(\psi^a_\mu)^{\cN=8} \ =& \ 
\exp\left[ - \tfrac{\ii}{2} \left(\theta+\pi\right) \Gamma_5 \right] \psi^a_\mu \, ,
\\[7pt]
\label{eq:TruncationDilatini}
(\chi^{abc})^{\cN=8} \ =& \ 
\exp\left[ - \tfrac{\ii}{2} \left(3\theta+\pi\right) \Gamma_5 \right] \epsilon^{abcd} \chi^d \, .
\end{align}
Every other field is set to zero. Here
\begin{equation}
\e^{2\ii\theta} \ = \ \frac{X+X^{-1}-\ii\varphi X}{X+X^{-1}+\ii\varphi X}
\end{equation}
is required in order to perform the reduction of the spinors (see also \cite{CSF}).

This truncation sends the bosonic part of the Lagrangian of the $\mc{N}=8$ theory into the Lorentzian action of the $Spin(4)$ model
\beq
\label{eq:ILorentz}
\begin{split}
	S_B \ &= \ \frac{1}{2 \kappa_4^2} \int \big[ R *1 - 2 X^{-2} \dd X \wedge * \dd X - \tfrac{1}{2} X^4 \dd \st \wedge * \dd \st + \gc^2 ( 8 + 2 X^2 + 2 \tilde{X}^2 ) * 1 \\
	& \qquad \quad  - \tfrac{1}{2} X^{-2} \big( \cF^I \wedge * \cF^I + \st X^2 \cF^I \wedge \cF^I \big) - \tfrac{1}{2} \tilde{X}^{-2} \big( \hat{\cF}^I \wedge * \hat{\cF}^I - \st X^2 \hat{\cF}^I \wedge \hat{\cF}^I \big) \big] \, .
\end{split}
\eeq
In the fermionic sector, we have four gravitini $\psi^a_\mu$ and four dilatini $\chi^a$, both transforming in the fundamental representation of $Spin(4)$. The fermionic action, which can be reconstructed from the supersymmetry variations and the bosonic action is, to lowest order in the fermions
\beq
\begin{split}
S_F = -\frac{1}{2\kappa^2_4} &\int \Big[ \bar{\psi}_\mu^a \Gamma^{\mu\nu\rho} \mathcal{D}_\nu \psi^a_\rho - \tfrac{1}{4\sqrt{2}} \eta^I_{ab} X^{-1} \cF^I_{\alpha\beta} \, \bar{\psi}_\mu^a \Gamma^{[\mu |} \Gamma^{\alpha\beta} \Gamma^{|\nu]} \psi_\nu^b \\
	&+ \tfrac{1}{4\sqrt{2}} \bar{\eta}^I_{ab} X^{-1} \tilde{X}^{-2} \hat{\cF}^I_{\alpha\beta} \, \bar{\psi}_\mu^a ( 1 + \ii \varphi X^2 \Gamma_5 ) \Gamma^{[\mu |} \Gamma^{\alpha\beta} \Gamma^{|\nu]} \psi_\nu^b  \\
	&+ \tfrac{\ii}{4} X^2 \partial_\nu \varphi \, \bar{\psi}_\mu^a \Gamma^{\mu\nu\rho} \Gamma_5 \psi_\rho^a + \tfrac{1}{\sqrt{2}} \g \, \bar{\psi}_\mu^a \left[ ( X + X^{-1} ) + \ii \j X \, \Gamma_5 \right]\Gamma^{\mu\nu} \psi_\nu^a  \\
	&+ \bar{\chi}^a \Gamma^\mu \mathcal{D}_\mu \chi^a + \tfrac{3\ii}{4} X^2 \partial_\mu \varphi \, \bar{\chi}^a \Gamma^\mu \Gamma_5 \chi^a  \\
	&- \sqrt{2} X^{-1} \partial_\nu X \, \bar{\psi}_\mu^a \Gamma^\nu \Gamma^\mu \chi^a +  \tfrac{\ii}{\sqrt{2}} X^2 \partial_\nu \varphi \, \bar{\psi}_\mu^a \Gamma_5 \Gamma^\nu \Gamma^\mu \chi^a  \\
	&+ \tfrac{1}{4} \eta^I_{ab} X^{-1} \cF^I_{\nu\rho} \, \bar{\psi}_\mu^a \Gamma^{\nu\rho} \Gamma^\mu \chi^b  + \tfrac{1}{4} \bar{\eta}^I_{ab} X^{-1} \tilde{X}^{-2} \hat{\cF}^I_{\nu\rho} \, \bar{\psi}_\mu^a ( 1 - \ii \varphi X^2 \Gamma_5 ) \Gamma^{\nu\rho} \Gamma^\mu \chi^b  \\
	&+ \g \, \bar{\psi}_\mu^a \left[ ( X - X^{-1} ) + \ii \j X \, \Gamma_5 \right] \Gamma^\mu \chi^a  \Big] \, \vol_G \, .
\end{split}
\eeq
The action is invariant under the following supersymmetry transformations with Majorana spinor parameter $\epsilon^a$:
\begin{align}
\label{eq:BulkFrame}
\delta \e^{\overline{\mu}}_\mu \ =& \ \tfrac{1}{2} \overline{\epsilon}^a\Gamma^{\overline{\mu}}\psi^a_\mu \, , \\[5pt]
\label{eq:BulkScalar}
\delta X \ =& \ \tfrac{1}{2\sqrt{2}} X \overline{\epsilon}^a\chi^a \, , \\[5pt]
\label{eq:BulkPseudoScalar}
\delta \j \ =& \ - \tfrac{\ii}{\sqrt{2}} X^{-2}\overline{\epsilon}^a\Gamma_5\chi^a \, , \\[5pt]
\label{eq:BulkGauge}
\delta \cA^I_\mu \ =& \ \tfrac{1}{\sqrt{2}} X \eta^I_{ab} \left( \overline{\epsilon}^a\psi^b_\mu - \tfrac{1}{\sqrt{2}} \overline{\epsilon}^a \Gamma_\mu \chi^b \right) \, , \\[5pt]
\label{eq:BulkGaugeHatted}
\delta \hat{\cA}^I_\mu \ =& \ - \tfrac{1}{\sqrt{2}} X \overline{\eta}^I_{ab} \left[ \overline{\epsilon}^a \left( X^{-2} + \ii \j \Gamma_5 \right) \psi^b_\mu + \tfrac{1}{\sqrt{2}} \overline{\epsilon}^a \Gamma_\mu \left( X^{-2} + \ii \j \Gamma_5 \right)  \chi^b \right] \, , \\[5pt]
	\begin{split}
	\delta \psi_\mu^a \ =& \ \mathcal{D}_\mu \epsilon^a - \tfrac{1}{8\sqrt{2}} \eta^I_{ab} X^{-1} \cF^I_{\nu\lambda} \Gamma^{\nu\lambda} \Gamma_\mu \epsilon^b + \tfrac{1}{8\sqrt{2}} \bar{\eta}^I_{ab} X^{-1} \tilde{X}^{-2} \hat{\cF}^I_{\nu\lambda} \Gamma^{\nu\lambda} \Gamma_\mu \big[ 1 + \ii \j X^2 \Gamma_5 \big] \epsilon^b \\
	& \ + \tfrac{\ii}{4} X^2 \partial_\mu \j \Gamma_5 \epsilon^a - \tfrac{1}{2\sqrt{2}} \g \big[ ( X + X^{-1} )  - \ii \j X \Gamma_5 \big] \Gamma_\mu \epsilon^a \, , 
	\end{split}\\[5pt]
	\begin{split}
	\delta \chi^a \ =& \ \tfrac{1}{8} \eta^I_{ab} X^{-1} \cF^I_{\nu\lambda} \Gamma^{\nu\lambda} \epsilon^b + \tfrac{1}{8} \bar{\eta}^I_{ab} X^{-1} \tilde{X}^{-2} \hat{\cF}^I_{\nu\lambda} \big[ 1 - \ii \j X^2 \Gamma_5 \big] \Gamma^{\nu\lambda} \epsilon^b \\
	&\ + \tfrac{1}{\sqrt{2}} \big[ X^{-1} \partial_\nu X + \tfrac{\ii}{2} X^2 \partial_\nu \j \Gamma_5 \big] \Gamma^\nu \epsilon^a + \tfrac{1}{2} \g \big[ ( X - X^{-1} ) + \ii \j X \Gamma_5 \big] \epsilon^a \, .
	\end{split}
\end{align}
These can be obtained from the $\mc{N}=8$ supersymmetry variations \cite[(3.1)--(3.5), (5.18), (5.19)]{deWit:1982bul} using \eqref{eq:TruncationGauge}--\eqref{eq:TruncationDilatini} together with the transformation of the $\mc{N}=8$ supersymmetry parameter
\beq
\label{eq:TruncSUSY}
(\epsilon^a)^{\cN=8} \ = \ \tfrac{1}{2} \exp\left[ - \tfrac{\ii}{2} \left( \theta + \pi \right) \Gamma_5 \right] \epsilon^a \, .
\eeq
For completeness, we record the equations of motion for the fermions. For the gravitino,  we find
\beq
\begin{split}
0 \ =& \ \Gamma^{\sigma\mu\nu} \hat{\mathcal{D}}_\mu \psi^a_{\nu} - \tfrac{1}{\sqrt{2}} \left( X^{-1} \partial_\mu X - \tfrac{\ii}{2} X^2 \partial_\mu \varphi \Gamma_5 \right) \Gamma^\mu \Gamma^\sigma \chi^a \\
	&+ \tfrac{1}{8} X^{-1} \left(   \eta^I_{ab}\cF^I_{\nu\rho} + \tilde{X}^{-2} \bar{\eta}^I_{ab}\hat{\cF}^I_{\nu\rho} ( 1 - \ii \varphi X^2 \Gamma_5 ) \right) \Gamma^{\nu\rho} \Gamma^\sigma \chi^b  \\
	&+ \tfrac{1}{2} \g \left[ ( X - X^{-1} ) + \ii \j X \Gamma_5 \right] \Gamma^\sigma \chi^a  \, ,
\end{split}
\eeq
where the supercovariant derivative is
\beq
\begin{split}
	\Gamma^{\sigma\mu\nu} \hat{\mathcal{D}}_\mu \psi^a_{\nu}  \ =& \ \Gamma^{\sigma\mu\nu} \mathcal{D}_\mu \psi^a_\nu - \tfrac{1}{2\sqrt{2}} \eta^I_{ab} X^{-1} \cF^I_{\rho\lambda} \left( \tfrac{1}{2} \Gamma^{\sigma\mu\rho\lambda} + G^{\sigma\rho} G^{\mu\lambda} \right) \psi_\mu^b  \\
	&+ \tfrac{1}{2\sqrt{2}} \bar{\eta}^I_{ab} X^{-1} \tilde{X}^{-2}  \hat{\cF}^I_{\rho\lambda} ( 1 + \ii \varphi X^2 \Gamma_5 ) \left( \tfrac{1}{2} \Gamma^{\sigma\mu\rho\lambda} + G^{\sigma\rho} G^{\mu\lambda} \right) \psi_\mu^b  \\
	&- \tfrac{\ii}{4} X^2 \partial_\nu \varphi \Gamma^{\sigma\mu\nu} \Gamma_5 \psi_\mu^a + \tfrac{1}{\sqrt{2}} \g \left[ (X + X^{-1})  + \ii \j X \Gamma_5 \right] \Gamma^{\sigma\mu} \psi_\mu^a  \, .
\end{split}
\eeq
For the dilatino, instead, we have
\beq
\label{eq:BulkDilatinoEOM}
0 \ = \ \Gamma^\mu \hat{\cD}_\mu \chi^a + \tfrac{3\ii}{4} X^2 \partial_\mu \j \Gamma^\mu \Gamma_5 \chi^a \, ,
\eeq
with supercovariant derivative
\beq
\begin{split}
\Gamma^\mu \cD_\mu \chi^a \ =& \ \Gamma^\mu \cD_\mu \chi^a - \tfrac{1}{8} \Gamma^\mu \left( \eta^I_{ab} X^{-1} \cF^I_{\nu\lambda} \Gamma^{\nu\lambda}  +  X^{-1} \tilde{X}^{-2} \bar{\eta}^I_{ab} \hat{\cF}^I_{\nu\lambda} \Gamma^{\nu\lambda} \left[ 1 - \ii \j X^2 \Gamma_5 \right] \right) \psi^b_\mu \\
& - \Gamma^\mu \left( \tfrac{1}{\sqrt{2}} X^{-1} \partial_\nu X \Gamma^\nu - \tfrac{\ii}{2\sqrt{2}} X^2 \partial_\nu \varphi \Gamma^\nu \Gamma_5 + \tfrac{1}{2} \g \left[ ( X - X^{-1} ) + \ii \j X \Gamma_5 \right] \right) \psi_\mu^a \, .
\end{split}
\eeq

Focusing on the scalar sector, we could choose to combine $X$ and $\j$ into a complex field by introducing $\tau \equiv \j + \ii X^{-2}$ and
\beq
z \equiv \frac{1+\ii \tau}{1-\ii \tau} \, .
\eeq
Then, the part of the action \eqref{eq:ILorentz} involving the scalar fields becomes
\beq
I_{\rm scalar} \ = \ \frac{1}{\kappa^2_4}\int \left[ - \frac{1}{\left( 1-\abs{z}^2 \right)^2}\partial_\mu z \partial^\mu \zbar + 2\g^2 \frac{3-\abs{z}^2}{1-\abs{z}^2} \right]*1 \, .
\eeq
In this formulation, it is easier to highlight features of the model expressed in $\cN=1$ language. The action for the scalar in a single $\mc{N}=1$ chiral multiplet has the form
\beq
I_{\rm chiral} \ = \ \frac{1}{\kappa^2_4}\int \left[ - \partial \overline{\partial} \mc{K} \, \partial_\mu z \partial^\mu \zbar - \mc{V} \right]*1 \, ,
\eeq
where $\mc{K}(z,\zbar)$ is the K\"ahler potential for the metric in the non-linear $\sigma$ model with metric $u \equiv \partial \overline{\partial} \mc{K}$, and $\mc{V}(z,\zbar)$ is the scalar potential, which can be obtained from the K\"ahler potential and the holomorphic superpotential $W(z)$ via the relation
\beq
\mc{V} \ = \ 4u^{-1} \, \partial \mc{W} \overline{\partial} \mc{W} - 3 \mc{W}^2 \, , \qquad \mc{W} \  \equiv \ \e^{\mc{K}/2}\abs{W} \, .
\eeq
Indeed, we immediately see that in our case $z$ parametrizes the Poincaré disc and the scalar kinetic term is derived as the $\sigma$-model with K\"ahler potential
\beq
\mc{K}(z,\zbar) \ = \ - \log \left( 1 - \abs{z}^2 \right) \, ,
\eeq
and the holomorphic superpotential $W(z)$ is just a constant, $\abs{W}^2 = 2\g^2$.
For completeness, we write down the expression for the superpotential in terms of the $X, \tilde{X}$ variables:
\beq
\label{eq:Superpotential}
\mc{W} \ = \ \sqrt{\frac{\g^2}{2}}\sqrt{2 + X^2 + \tilde{X}^2} \, ,
\eeq
which appears in Section \ref{subsec:HoloRen}, when we perform the holographic renormalization.

\medskip

In Wick rotating to Riemannian signature, we cannot define Majorana spinors and impose reality conditions on the fields. Thus, we relax the constraint and we consider Dirac spinors and complex scalar fields. In Section \ref{subsec:OSAction} we impose additional reality conditions for a truncated subsector of the theory.

\medskip

The first upshot of reducing from $\mc{N}=8$ supergravity is the possibility of appealing to the results of \cite{FreedmanPufuEtAl} to justify the supersymmetry-preserving counterterm \eqref{eq:properCT}. In order to match their conventions, we write the supersymmetries of the $\mc{N}=4$ model as $\epsilon^i$ with $i=1, \dots, 8$, constrained by
\beq
\Pi^i_{\ph{i}j}\epsilon^j \ = \ \epsilon^i \, , \qquad \Pi^i_{\ph{i}j}\Pi^j_{\ph{j}k} \ = \ \Pi^i_{\ph{i}k} \, , \qquad \Pi^i_{\ph{i}i} \ = \ 4 \, ,
\eeq
where the non-vanishing components are $\Pi^a_{\ph{a}b} = \delta^a_b$. As we showed, the reduction outlined in \eqref{eq:TruncationGauge}--\eqref{eq:TruncationDilatini} is compatible with the equations of motion of the gravitini, sending those of the $\cN=8$ model to those of the $\cN=4$ model. Furthermore, taking the definition of the symmetric tensor $A_1^{ij}$ from \cite{deWit:1982bul}
\beq
A_1^{ij} \ \equiv \ \frac{4}{21} \left( u^{kj}_{\ph{kj}IJ} + v^{kjIJ} \right) \left( u_{km}^{\ph{km}JK} u^{im}_{\ph{im}KI} - v_{kmJK}v^{imKI} \right)
\eeq
and applying \eqref{eq:TruncationScalaru}, \eqref{eq:TruncationScalarv} gives
\beq
\begin{split}
A_1^{ab} \ &= \ \frac{1}{2} \left(X + X^{-1}+\ii\varphi X\right)\e^{\ii\theta} \, \delta^{ab} \, , \qquad A_1^{\overline{ab}} \ = \ \frac{1}{2} \left(X + X^{-1}+\ii\varphi X\right)\e^{\ii\theta} \, \delta^{\overline{ab}} \, .
\end{split}
\eeq
In Appendix C of \cite{FreedmanPufuEtAl} it is then shown that for a truncation of $\mc{N}=8$ obtained analogously to ours, the supersymmetry-preserving counterterm has the form
\begin{equation}
\begin{split}
S_{\rm ct,s}^{\rm SUSY} \ &= \ - \frac{2\sqrt{2}\g }{p}\frac{1}{\kappa^2_4}
\int_{M_\delta}
\Tr\sqrt{\Pi A_1A_1^\dagger} \, \vol_h \\
&= \ - \sqrt{2}\g \frac{1}{\kappa^2_4} \int_{M_\delta} \sqrt{(X+X^{-1})^2 + \j^2 X^2} \, \vol_h \, ,
\end{split}
\end{equation}
which matches our \eqref{eq:properCT}.

These results are based on the requirement that the on-shell action for the ``kink Ansatz'' metric has a vanishing contribution from the conformal boundary, as required by supersymmetry. We have also separately verified this by applying the arguments in Section 6 of \cite{FreedmanPufuEtAl} to our theory \eqref{eq:ILorentz} (setting the gauge fields to zero): write the on-shell action as a sum of squares of objects implementing the BPS conditions, and then impose the vanishing of the boundary contribution.

\section{Boundary supersymmetry}
\label{app:BdrySUSY}

The gravitational free energy, corresponding in our case to the Legendre transform of the renormalized gravitational on-shell action \eqref{eq:LegendreTransformation}, should be interpreted as a functional of the boundary data representing sources of the field theory. As such, it is only supersymmetric provided that these are transformed into each other under a supersymmetry variation. This point has been stressed in e.g.\ \cite{Amsel:2008iz, FreedmanPufu, FreedmanPufuEtAl}. In this appendix, we show that this holds by expanding the bulk supersymmetry transformations \eqref{eq:BulkFrame}--\eqref{eq:BulkGauge} and showing that we match the supersymmetry variations of the three-dimensional off-shell conformal supergravity of \cite{Banerjee:2015uee}. These supplement the considerations in Section \ref{subsec:ExpansionSUSY} to provide a more complete picture. Being interested in the global limit of the supersymmetry, we consider backgrounds in which the variation of the gravitino \eqref{eq:BulkGravitino} has been set to zero and we consistently set the bulk gravitino itself to zero.

\medskip

We choose the basis (in Lorentzian signature)
\beq
\begin{aligned}
\Gamma_{\overline{0}} &= \begin{pmatrix}
0 & \ii \sigma_1 \\
\ii \sigma_1 & 0
\end{pmatrix} \, , &\qquad
\Gamma_{\overline{1}} &= \begin{pmatrix}
0 & \sigma_2 \\
\sigma_2 & 0
\end{pmatrix} \, , &\qquad 
\Gamma_{\overline{2}} = \Gamma_{\overline{z}} &= \begin{pmatrix}
\identity_2 & 0 \\ 
0 & - \identity_2
\end{pmatrix} \, , \\
\Gamma_{\overline{3}} &= \begin{pmatrix}
0 & \sigma_3 \\
\sigma_3 & 0
\end{pmatrix} \, , &\qquad
\Gamma_5 &= \begin{pmatrix}
0 & - \ii \identity_2 \\
\ii \identity_2 & 0
\end{pmatrix} \, , &\qquad
\mathscr{C}_4 &= \begin{pmatrix}
0 & \ii \sigma_2 \\
\ii \sigma_2 & 0
\end{pmatrix} \, ,
\end{aligned}
\eeq
and decompose them as
\beq
\Gamma_{\overline{0}} \ = \ \sigma_1 \otimes \gamma_{\overline{0}} \, , \qquad \Gamma_{\overline{1}} \ = \ \sigma_1 \otimes \gamma_{\overline{1}} \, , \qquad \Gamma_{\overline{3}} \ = \ \sigma_1 \otimes \gamma_{\overline{2}} \, , \qquad \mathscr{C}_4 \ = \ \sigma_1 \otimes \mathscr{C}_3 \, ,
\eeq
where $\gamma_{\overline{i}}$ is a basis of Cliff$(2,1)$ and $\mathscr{C}_3$ is the three-dimensional charge conjugation matrix.
Notice that the Majorana condition in four dimensions reduces to the Majorana condition in three dimensions.
As in the bulk of the text, we use $L$ and $R$ to indicate the chirality with respect to $\Gamma_{\overline{z}}$.

\medskip

Near the boundary, the bulk bosonic fields can be expanded as \eqref{Xexp}--\eqref{hataIexp}, and the bulk supersymmetry parameter has the form
\beq
\epsilon^a \ = \ z^{-1/2} \begin{pmatrix}
\varepsilon^a_L \\ 0
\end{pmatrix}  + z^{1/2}  \begin{pmatrix}
0 \\ \xi^a_R
\end{pmatrix} + o(z) \, .
\eeq 
For the dilatino, expanding \eqref{eq:BulkDilatinoEOM} near the boundary gives
\beq
\chi^a \ = \ z^{3/2} \, \chi^a_0 + z^{5/2} \, \chi_1^a + o(z^2) \, ,
\eeq
with
\beq
\chi^a_{1L} \ = \ - \slashed{D}\chi^a_{0R} - \tfrac{3}{4}\j_1 \, \chi^a_{0R} \, , \qquad \qquad \chi^a_{1R} \ = \ \slashed{D}\chi^a_{0L} + \tfrac{3}{4}\j_1 \, \chi^a_{0L} \, .
\eeq
These expansions leads to the following supersymmetry variations of the scalars and gauge fields
\beq
\label{eq:LeadingBulkScalars}
\begin{alignedat}{2}
\delta X_1 \ &= \ \frac{1}{2\sqrt{2}}\overline{\varepsilon}^a_L \chi^a_{0R} \, , \qquad & \delta X_2 \ &= \ \frac{1}{2\sqrt{2}} \left( \overline{\varepsilon}^a_L \chi^a_{1R} + \overline{\xi}^a_R \chi^a_{0L} + X_1 \overline{\varepsilon}^a_L \chi^a_{0R} \right) \, , \\
\delta \j_1 \ &= \ \frac{1}{\sqrt{2}} \overline{\varepsilon}^a_L \chi^a_{0L} \, , \qquad & \delta \j_2 \ &= \ \frac{1}{\sqrt{2}} \left( \overline{\varepsilon}^a_L \chi^a_{1L} - \overline{\xi}^a_R \chi^a_{0R} - 2X_1 \overline{\varepsilon}^a_L \chi^a_{0L} \right) \, , \\
\delta A_i \ &= \ - \frac{1}{2}\overline{\varepsilon}^a_L\gamma_i \chi^a_{0L} \, , \qquad & \delta \hat{A}_i \ &= \ -\frac{1}{2}\overline{\varepsilon}^a\gamma_i \chi^a_{0L} \, .
\end{alignedat}
\eeq
Recall that in Section \ref{subsec:ExpansionSUSY} we identified the leading order gravitino $\chi^a_{0L}$ with the gravitino at the boundary, so $\j_1$ is indeed transformed by supersymmetry into another source. However, this is not true for $X_1$, and indeed we know that this should not hold. Since the dual operator to $X$ has scaling dimension $1$, its source cannot be $X_1$, and we identified it in \eqref{eq:Identification2} with $\Xsource$. In fact
\beq
\begin{split}
-\kappa^2_4 \delta \Xsource \ &= \ - \left( 2 X_1 \delta X_1 - 2 \delta X_2 + \j_1 \delta \j_1 \right) \\
&= \ \frac{1}{\sqrt{2}} \left( \overline{\varepsilon}^a_L \slashed{D} \chi^a_{0L} + \overline{\xi}^a_R\chi^a_{0L} - \tfrac{1}{4}\j_1 \, \overline{\varepsilon}^a_L\chi^a_{0L} \right) \, ,
\end{split}
\eeq
and using \eqref{eq:Identification1} and \eqref{eq:Identification2} we find
\beq
- \kappa^2_4 \delta \Xsource \ = \ \frac{1}{2} \left( \overline{\tredSUGRAQ}^a \slashed{D}(\chi^a)^{3d} + \overline{\tredSUGRAS}^a(\chi^a)^{3d} \right) \, .
\eeq
Notice that not only does $\delta \Xsource$ only contain the fermionic source, but it also corresponds --- as it should --- to the supersymmetry variation of the corresponding boundary scalar in $3d$ off-shell conformal supergravity, as do the variations of $\j_1$ and the gauge fields \cite{Banerjee:2015uee}
\beq
\label{eq:BdryScalarsOffShell}
\begin{aligned}
\delta S_1 \ &= \ \frac{1}{4} \overline{\tredSUGRAQ}^a (\chi^a)^{3d} \, , &\qquad
\delta S_2 \ &= \ \frac{1}{2} \overline{\tredSUGRAQ}^a \slashed{D} (\chi^a)^{3d} + \frac{1}{2} \overline{\tredSUGRAS}^a (\chi^a)^{3d} \, , \\
\delta A^I_i \ &= \ - \frac{1}{2\sqrt{2}}\eta^I_{ab}\overline{\zeta}^a \gamma_i (\chi^a)^{3d} \, , &\qquad
\delta \hat{A}^I_i \ &= \ - \frac{1}{2\sqrt{2}}\overline{\eta}^I_{ab}\overline{\zeta}^a \gamma_i (\chi^a)^{3d} \, .
\end{aligned}
\eeq
Showing that the bulk supersymmetry variations match those of the expected boundary conformal supergravity, provided we identify correctly the sources for the dual boundary operators, provides additional evidence to the holographic renormalization procedure described in Section \ref{subsec:HoloRen}, including the necessity of the supersymmetry-preserving counterterm and the Legendre transformation.

\section{Holographic supersymmetric Ward identity}
\label{app:SUSYWard}

In Section \ref{subsec:HoloRen} we looked at the holographic Ward identities corresponding to boundary Weyl symmetry, R-symmetry, and diffeomorphism invariance. However, when we include non-vanishing fermionic fields, we may also consider the Ward identities corresponding to the boundary $\mc{Q}$ and $\mc{S}$ supersymmetry of the background conformal supergravity. Here, we focus on a specific one that appears as the constraint \eqref{eq:BdryDilatino2} in the bulk supersymmetry equations.

Generically, a variation of the gravitational on-shell action under a change of the boundary data can be written as
\beq
\label{eq:VariationSpinors}
\begin{split}
\delta \tilde{\mathbb{S}} \ &= \ \int_{M_3} \Big[ \ \delta \e^{\overline{i}}_i \, \langle {T}_{\overline{i}}^{i} \rangle + \delta \overline{\psi}^a_i \langle \mc{S}^{ai} \rangle + \delta \j_1 \langle \mc{O}_{\Delta_+} \rangle + \delta A^I_i \langle \mathscr{J}^{Ii} \rangle  + \delta \hat{A}^I_i \langle \hat{\mathscr{J}}^{Ii} \rangle \\
& \qquad \qquad  + \delta \overline{\chi}^a \langle \lambda^a \rangle + \delta \Xsource \langle \mc{O}_{\Delta_-} \rangle \ \Big] \vol_g
\end{split}
\eeq
which generalizes \eqref{eq:Variation} to include the fermionic boundary sources and VEV written in \eqref{eq:3dWeylmultiplet} and \eqref{eq:3dTmultiplet}. Here, we focus on the supersymmetric Ward identity in a supersymmetric classical background with vanishing gravitino using the supersymmetry transformations described in Appendix \ref{app:BdrySUSY}. The gravitational free energy is supersymmetric provided
\beq
\begin{split}
0 \ &= \ \frac{1}{2\kappa^2_4} \bigg\{  \partial_{{i}} \langle \mc{O}_{\Delta_-} \rangle  \overline{\zeta}_a \gamma^{{i}}(\chi^a)^{3d} + \kappa^2_4 \langle \mc{O}_{\Delta_+} \rangle \overline{\zeta}_a(\chi^a)^{3d} + 2 \langle \mc{O}_{\Delta_-} \rangle \overline{\vartheta}^a(\chi^a)^{3d}  \\
& \qquad \qquad \qquad - \tfrac{\kappa^2_4}{\sqrt{2}} \left( \eta^I_{ab} \langle \mathscr{J}^I_i \rangle +  \overline{\eta}^I_{ab} \langle \hat{\mathscr{J}}^I_i \rangle \right) \overline{\zeta}^a \gamma^i (\chi^b)^{3d} \bigg\}
\end{split}
\eeq
or equivalently
\beq
\begin{split}
0 \ &= \ \frac{1}{2\kappa^2_4} \, \overline{(\chi^a)^{3d}} \bigg\{  \kappa^2_4 \langle \mc{O}_{\Delta_+} \rangle \zeta^a - \partial_{{i}} \langle \mc{O}_{\Delta_-} \rangle \gamma^{{i}}\zeta^a + 2 \langle \mc{O}_{\Delta_-} \rangle {\vartheta}^a \\
& \qquad \qquad \qquad + \tfrac{\kappa^2_4}{\sqrt{2}} \left( \eta^I_{ab} \langle \mathscr{J}^I_i \rangle + \overline{\eta}^I_{ab} \langle \hat{\mathscr{J}}^I_i \rangle \right) \gamma^i \zeta^b \bigg\} \, ,
\end{split}
\eeq
which by the arbitrariness in the choice of the gravitino gives \eqref{eq:BdryDilatino2}.

\bibliographystyle{./SUSYCountertermsFiles/JHEP}
{\small
\bibliography{./SUSYCountertermsFiles/Bib_SUSY_Counterterms}

\providecommand{\href}[2]{#2}\begingroup\raggedright\begin{thebibliography}{10}

\bibitem{Closset:2012ru}
C.~Closset, T.~T. Dumitrescu, G.~Festuccia and Z.~Komargodski,
  \emph{{Supersymmetric Field Theories on Three-Manifolds}},
  \href{http://dx.doi.org/10.1007/JHEP05(2013)017}{\emph{JHEP} {\bf 05} (2013)
  017}, [\href{http://arxiv.org/abs/1212.3388}{{\tt 1212.3388}}].

\bibitem{Witten:1988ze}
E.~Witten, \emph{{Topological Quantum Field Theory}},
  \href{http://dx.doi.org/10.1007/BF01223371}{\emph{Commun. Math. Phys.} {\bf
  117} (1988) 353}.

\bibitem{BenettiGenolini:2017zmu}
P.~Benetti~Genolini, P.~Richmond and J.~Sparks, \emph{{Topological AdS/CFT}},
  \href{http://dx.doi.org/10.1007/JHEP12(2017)039}{\emph{JHEP} {\bf 12} (2017)
  039}, [\href{http://arxiv.org/abs/1707.08575}{{\tt 1707.08575}}].

\bibitem{GRS1}
P.~Benetti~Genolini, P.~Richmond and J.~Sparks, \emph{{Gravitational free
  energy in topological AdS/CFT}},
  \href{http://dx.doi.org/10.1007/JHEP09(2018)100}{\emph{JHEP} {\bf 09} (2018)
  100}, [\href{http://arxiv.org/abs/1804.08625}{{\tt 1804.08625}}].

\bibitem{Bobev:2019ylk}
N.~Bobev, F.~F. Gautason and K.~Hristov, \emph{{Holographic dual of the
  $\Omega$ -background}},
  \href{http://dx.doi.org/10.1103/PhysRevD.100.021901}{\emph{Phys. Rev. D} {\bf
  100} (2019) 021901}, [\href{http://arxiv.org/abs/1903.05095}{{\tt
  1903.05095}}].

\bibitem{BenettiGenolini:2019wxg}
P.~Benetti~Genolini and P.~Richmond, \emph{{Topological AdS/CFT and the
  $\Omega$ deformation}},
  \href{http://dx.doi.org/10.1007/JHEP10(2019)115}{\emph{JHEP} {\bf 10} (2019)
  115}, [\href{http://arxiv.org/abs/1907.12561}{{\tt 1907.12561}}].

\bibitem{Costello:2016mgj}
K.~Costello and S.~Li, \emph{{Twisted supergravity and its quantization}},
  \href{http://arxiv.org/abs/1606.00365}{{\tt 1606.00365}}.

\bibitem{Bonetti:2016nma}
F.~Bonetti and L.~Rastelli, \emph{{Supersymmetric localization in AdS$_{5}$ and
  the protected chiral algebra}},
  \href{http://dx.doi.org/10.1007/JHEP08(2018)098}{\emph{JHEP} {\bf 08} (2018)
  098}, [\href{http://arxiv.org/abs/1612.06514}{{\tt 1612.06514}}].

\bibitem{Mezei:2017kmw}
M.~Mezei, S.~S. Pufu and Y.~Wang, \emph{{A 2d/1d Holographic Duality}},
  \href{http://arxiv.org/abs/1703.08749}{{\tt 1703.08749}}.

\bibitem{Costello:2017fbo}
K.~Costello, \emph{{Holography and Koszul duality: the example of the $M2$
  brane}},  \href{http://arxiv.org/abs/1705.02500}{{\tt 1705.02500}}.

\bibitem{Costello:2018zrm}
K.~Costello and D.~Gaiotto, \emph{{Twisted Holography}},
  \href{http://arxiv.org/abs/1812.09257}{{\tt 1812.09257}}.

\bibitem{Gaiotto:2019wcc}
D.~Gaiotto and J.~Oh, \emph{{Aspects of $\Omega$-deformed M-theory}},
  \href{http://arxiv.org/abs/1907.06495}{{\tt 1907.06495}}.

\bibitem{Costello:2020jbh}
K.~Costello and N.~M. Paquette, \emph{{Twisted Supergravity and Koszul Duality:
  A case study in AdS$_3$}},  \href{http://arxiv.org/abs/2001.02177}{{\tt
  2001.02177}}.

\bibitem{Oh:2020hph}
J.~Oh and Y.~Zhou, \emph{{Feynman diagrams and $\Omega$-deformed M-theory}},
  \href{http://arxiv.org/abs/2002.07343}{{\tt 2002.07343}}.

\bibitem{Gaiotto:2020vqj}
D.~Gaiotto and J.~Abajian, \emph{{Twisted M2 brane holography and sphere
  correlation functions}},  \href{http://arxiv.org/abs/2004.13810}{{\tt
  2004.13810}}.

\bibitem{Li:2019qzx}
S.~Li and J.~Troost, \emph{{Pure and Twisted Holography}},
  \href{http://dx.doi.org/10.1007/JHEP03(2020)144}{\emph{JHEP} {\bf 03} (2020)
  144}, [\href{http://arxiv.org/abs/1911.06019}{{\tt 1911.06019}}].

\bibitem{Li:2020nei}
S.~Li and J.~Troost, \emph{{Twisted String Theory in Anti-de Sitter Space}},
  \href{http://arxiv.org/abs/2005.13817}{{\tt 2005.13817}}.

\bibitem{Dabholkar:2014wpa}
A.~Dabholkar, N.~Drukker and J.~Gomes, \emph{{Localization in supergravity and
  quantum $AdS_4/CFT_3$ holography}},
  \href{http://dx.doi.org/10.1007/JHEP10(2014)090}{\emph{JHEP} {\bf 10} (2014)
  090}, [\href{http://arxiv.org/abs/1406.0505}{{\tt 1406.0505}}].

\bibitem{Bae:2015eoa}
J.~Bae, C.~Imbimbo, S.-J. Rey and D.~Rosa, \emph{{New Supersymmetric
  Localizations from Topological Gravity}},
  \href{http://dx.doi.org/10.1007/JHEP03(2016)169}{\emph{JHEP} {\bf 03} (2016)
  169}, [\href{http://arxiv.org/abs/1510.00006}{{\tt 1510.00006}}].

\bibitem{Imbimbo:2018duh}
C.~Imbimbo and D.~Rosa, \emph{{The topological structure of supergravity: an
  application to supersymmetric localization}},
  \href{http://dx.doi.org/10.1007/JHEP05(2018)112}{\emph{JHEP} {\bf 05} (2018)
  112}, [\href{http://arxiv.org/abs/1801.04940}{{\tt 1801.04940}}].

\bibitem{Hristov:2018lod}
K.~Hristov, I.~Lodato and V.~Reys, \emph{{On the quantum entropy function in 4d
  gauged supergravity}},
  \href{http://dx.doi.org/10.1007/JHEP07(2018)072}{\emph{JHEP} {\bf 07} (2018)
  072}, [\href{http://arxiv.org/abs/1803.05920}{{\tt 1803.05920}}].

\bibitem{deWit:2018dix}
B.~de~Wit, S.~Murthy and V.~Reys, \emph{{BRST quantization and equivariant
  cohomology: localization with asymptotic boundaries}},
  \href{http://dx.doi.org/10.1007/JHEP09(2018)084}{\emph{JHEP} {\bf 09} (2018)
  084}, [\href{http://arxiv.org/abs/1806.03690}{{\tt 1806.03690}}].

\bibitem{Jeon:2018kec}
I.~Jeon and S.~Murthy, \emph{{Twisting and localization in supergravity:
  equivariant cohomology of BPS black holes}},
  \href{http://dx.doi.org/10.1007/JHEP03(2019)140}{\emph{JHEP} {\bf 03} (2019)
  140}, [\href{http://arxiv.org/abs/1806.04479}{{\tt 1806.04479}}].

\bibitem{BenettiGenolini:2019jdz}
P.~Benetti~Genolini, J.~M. Perez Ipi\~na and J.~Sparks, \emph{{Localization of
  the action in AdS/CFT}},
  \href{http://dx.doi.org/10.1007/JHEP10(2019)252}{\emph{JHEP} {\bf 10} (2019)
  252}, [\href{http://arxiv.org/abs/1906.11249}{{\tt 1906.11249}}].

\bibitem{FreedmanPufu}
D.~Z. Freedman and S.~S. Pufu, \emph{{The holography of $F$-maximization}},
  \href{http://dx.doi.org/10.1007/JHEP03(2014)135}{\emph{JHEP} {\bf 03} (2014)
  135}, [\href{http://arxiv.org/abs/1302.7310}{{\tt 1302.7310}}].

\bibitem{FreedmanPufuEtAl}
D.~Z. Freedman, K.~Pilch, S.~S. Pufu and N.~P. Warner, \emph{{Boundary Terms
  and Three-Point Functions: An AdS/CFT Puzzle Resolved}},
  \href{http://dx.doi.org/10.1007/JHEP06(2017)053}{\emph{JHEP} {\bf 06} (2017)
  053}, [\href{http://arxiv.org/abs/1611.01888}{{\tt 1611.01888}}].

\bibitem{Genolini:2016ecx}
P.~Benetti~Genolini, D.~Cassani, D.~Martelli and J.~Sparks, \emph{{Holographic
  renormalization and supersymmetry}},
  \href{http://dx.doi.org/10.1007/JHEP02(2017)132}{\emph{JHEP} {\bf 02} (2017)
  132}, [\href{http://arxiv.org/abs/1612.06761}{{\tt 1612.06761}}].

\bibitem{Halmagyi:2017hmw}
N.~Halmagyi and S.~Lal, \emph{{On the on-shell: the action of AdS$_{4}$ black
  holes}}, \href{http://dx.doi.org/10.1007/JHEP03(2018)146}{\emph{JHEP} {\bf
  03} (2018) 146}, [\href{http://arxiv.org/abs/1710.09580}{{\tt 1710.09580}}].

\bibitem{Cabo-Bizet:2017xdr}
A.~Cabo-Bizet, U.~Kol, L.~A. Pando~Zayas, I.~Papadimitriou and V.~Rathee,
  \emph{{Entropy functional and the holographic attractor mechanism}},
  \href{http://dx.doi.org/10.1007/JHEP05(2018)155}{\emph{JHEP} {\bf 05} (2018)
  155}, [\href{http://arxiv.org/abs/1712.01849}{{\tt 1712.01849}}].

\bibitem{Breitenlohner:1982jf}
P.~Breitenlohner and D.~Z. Freedman, \emph{{Stability in Gauged Extended
  Supergravity}},
  \href{http://dx.doi.org/10.1016/0003-4916(82)90116-6}{\emph{Annals Phys.}
  {\bf 144} (1982) 249}.

\bibitem{Klebanov:1999tb}
I.~R. Klebanov and E.~Witten, \emph{{AdS / CFT correspondence and symmetry
  breaking}},
  \href{http://dx.doi.org/10.1016/S0550-3213(99)00387-9}{\emph{Nucl. Phys. B}
  {\bf 556} (1999) 89--114}, [\href{http://arxiv.org/abs/hep-th/9905104}{{\tt
  hep-th/9905104}}].

\bibitem{DFR}
A.~Das, M.~Fischler and M.~Ro\ifmmode~\check{c}\else \v{c}\fi{}ek,
  \emph{Super-higgs effect in a new class of scalar models and a model of super
  qed}, \href{http://dx.doi.org/10.1103/PhysRevD.16.3427}{\emph{Phys. Rev. D}
  {\bf 16} (Dec, 1977) 3427--3436}.

\bibitem{Vafa:1994tf}
C.~Vafa and E.~Witten, \emph{{A Strong coupling test of S duality}},
  \href{http://dx.doi.org/10.1016/0550-3213(94)90097-3}{\emph{Nucl. Phys. B}
  {\bf 431} (1994) 3--77}, [\href{http://arxiv.org/abs/hep-th/9408074}{{\tt
  hep-th/9408074}}].

\bibitem{Festuccia:2011ws}
G.~Festuccia and N.~Seiberg, \emph{{Rigid Supersymmetric Theories in Curved
  Superspace}}, \href{http://dx.doi.org/10.1007/JHEP06(2011)114}{\emph{JHEP}
  {\bf 06} (2011) 114}, [\href{http://arxiv.org/abs/1105.0689}{{\tt
  1105.0689}}].

\bibitem{Witten:1989sx}
E.~Witten, \emph{{Topology Changing Amplitudes in (2+1)-Dimensional Gravity}},
  \href{http://dx.doi.org/10.1016/0550-3213(89)90591-9}{\emph{Nucl. Phys. B}
  {\bf 323} (1989) 113--140}.

\bibitem{Blau:1991bn}
M.~Blau and G.~Thompson, \emph{{N=2 topological gauge theory, the Euler
  characteristic of moduli spaces, and the Casson invariant}},
  \href{http://dx.doi.org/10.1007/BF02097057}{\emph{Commun. Math. Phys.} {\bf
  152} (1993) 41--72}, [\href{http://arxiv.org/abs/hep-th/9112012}{{\tt
  hep-th/9112012}}].

\bibitem{Gaiotto:2008sd}
D.~Gaiotto and E.~Witten, \emph{{Janus Configurations, Chern-Simons Couplings,
  And The theta-Angle in N=4 Super Yang-Mills Theory}},
  \href{http://dx.doi.org/10.1007/JHEP06(2010)097}{\emph{JHEP} {\bf 06} (2010)
  097}, [\href{http://arxiv.org/abs/0804.2907}{{\tt 0804.2907}}].

\bibitem{Kapustin:2009cd}
A.~Kapustin and N.~Saulina, \emph{{Chern-Simons-Rozansky-Witten topological
  field theory}},
  \href{http://dx.doi.org/10.1016/j.nuclphysb.2009.07.006}{\emph{Nucl. Phys. B}
  {\bf 823} (2009) 403--427}, [\href{http://arxiv.org/abs/0904.1447}{{\tt
  0904.1447}}].

\bibitem{Koh:2009um}
E.~Koh, S.~Lee and S.~Lee, \emph{{Topological Chern-Simons Sigma Model}},
  \href{http://dx.doi.org/10.1088/1126-6708/2009/09/122}{\emph{JHEP} {\bf 09}
  (2009) 122}, [\href{http://arxiv.org/abs/0907.1641}{{\tt 0907.1641}}].

\bibitem{Mikhaylov:2015nsa}
V.~Mikhaylov, \emph{{Analytic Torsion, 3d Mirror Symmetry And Supergroup
  Chern-Simons Theories}},  \href{http://arxiv.org/abs/1505.03130}{{\tt
  1505.03130}}.

\bibitem{Cordova:2016emh}
C.~Cordova, T.~T. Dumitrescu and K.~Intriligator, \emph{{Multiplets of
  Superconformal Symmetry in Diverse Dimensions}},
  \href{http://dx.doi.org/10.1007/JHEP03(2019)163}{\emph{JHEP} {\bf 03} (2019)
  163}, [\href{http://arxiv.org/abs/1612.00809}{{\tt 1612.00809}}].

\bibitem{Dumitrescu:2016ltq}
T.~T. Dumitrescu, \emph{{An introduction to supersymmetric field theories in
  curved space}}, \href{http://dx.doi.org/10.1088/1751-8121/aa62f5}{\emph{J.
  Phys. A} {\bf 50} (2017) 443005},
  [\href{http://arxiv.org/abs/1608.02957}{{\tt 1608.02957}}].

\bibitem{Banerjee:2015uee}
N.~Banerjee, B.~de~Wit and S.~Katmadas, \emph{{The off-shell c-map}},
  \href{http://dx.doi.org/10.1007/JHEP01(2016)156}{\emph{JHEP} {\bf 01} (2016)
  156}, [\href{http://arxiv.org/abs/1512.06686}{{\tt 1512.06686}}].

\bibitem{Trunc}
M.~Cvetic, H.~Lu and C.~Pope, \emph{{Four-dimensional N=4, SO(4) gauged
  supergravity from D = 11}},
  \href{http://dx.doi.org/10.1016/S0550-3213(99)00828-7}{\emph{Nucl. Phys. B}
  {\bf 574} (2000) 761--781}, [\href{http://arxiv.org/abs/hep-th/9910252}{{\tt
  hep-th/9910252}}].

\bibitem{Lee:2008cr}
K.~Lee, S.~Lee and J.-H. Park, \emph{{Topological Twisting of Multiple M2-brane
  Theory}}, \href{http://dx.doi.org/10.1088/1126-6708/2008/11/014}{\emph{JHEP}
  {\bf 11} (2008) 014}, [\href{http://arxiv.org/abs/0809.2924}{{\tt
  0809.2924}}].

\bibitem{Aharony:2008ug}
O.~Aharony, O.~Bergman, D.~L. Jafferis and J.~Maldacena, \emph{{N=6
  superconformal Chern-Simons-matter theories, M2-branes and their gravity
  duals}}, \href{http://dx.doi.org/10.1088/1126-6708/2008/10/091}{\emph{JHEP}
  {\bf 10} (2008) 091}, [\href{http://arxiv.org/abs/0806.1218}{{\tt
  0806.1218}}].

\bibitem{Hosomichi:2008jb}
K.~Hosomichi, K.-M. Lee, S.~Lee, S.~Lee and J.~Park, \emph{{N=5,6
  Superconformal Chern-Simons Theories and M2-branes on Orbifolds}},
  \href{http://dx.doi.org/10.1088/1126-6708/2008/09/002}{\emph{JHEP} {\bf 09}
  (2008) 002}, [\href{http://arxiv.org/abs/0806.4977}{{\tt 0806.4977}}].

\bibitem{Fefferman:2007rka}
C.~Fefferman and C.~R. Graham, \emph{{The ambient metric}}, {\emph{Ann. Math.
  Stud.} {\bf 178} (2011) 1--128}, [\href{http://arxiv.org/abs/0710.0919}{{\tt
  0710.0919}}].

\bibitem{Emparan:1999pm}
R.~Emparan, C.~V. Johnson and R.~C. Myers, \emph{{Surface terms as counterterms
  in the AdS/CFT correspondence}},
  \href{http://dx.doi.org/10.1103/PhysRevD.60.104001}{\emph{Phys. Rev.} {\bf
  D60} (1999) 104001}, [\href{http://arxiv.org/abs/hep-th/9903238}{{\tt
  hep-th/9903238}}].

\bibitem{Taylor:2000xw}
M.~Taylor, \emph{{More on counterterms in the gravitational action and
  anomalies}},  \href{http://arxiv.org/abs/hep-th/0002125}{{\tt
  hep-th/0002125}}.

\bibitem{deHaro:2000vlm}
S.~de~Haro, S.~N. Solodukhin and K.~Skenderis, \emph{{Holographic
  reconstruction of space-time and renormalization in the AdS/CFT
  correspondence}},
  \href{http://dx.doi.org/10.1007/s002200100381}{\emph{Commun. Math. Phys.}
  {\bf 217} (2001) 595--622}, [\href{http://arxiv.org/abs/hep-th/0002230}{{\tt
  hep-th/0002230}}].

\bibitem{Gauntlett:2018vhk}
J.~P. Gauntlett and C.~Rosen, \emph{{Susy Q and spatially modulated
  deformations of ABJM theory}},
  \href{http://dx.doi.org/10.1007/JHEP10(2018)066}{\emph{JHEP} {\bf 10} (2018)
  066}, [\href{http://arxiv.org/abs/1808.02488}{{\tt 1808.02488}}].

\bibitem{Bobev:2018wbt}
N.~Bobev, V.~S. Min, K.~Pilch and F.~Rosso, \emph{{Mass Deformations of the
  ABJM Theory: The Holographic Free Energy}},
  \href{http://dx.doi.org/10.1007/JHEP03(2019)130}{\emph{JHEP} {\bf 03} (2019)
  130}, [\href{http://arxiv.org/abs/1812.01026}{{\tt 1812.01026}}].

\bibitem{Arav:2018njv}
I.~Arav, J.~P. Gauntlett, M.~Roberts and C.~Rosen, \emph{{Spatially modulated
  and supersymmetric deformations of ABJM theory}},
  \href{http://dx.doi.org/10.1007/JHEP04(2019)099}{\emph{JHEP} {\bf 04} (2019)
  099}, [\href{http://arxiv.org/abs/1812.11159}{{\tt 1812.11159}}].

\bibitem{Bobev:2020pjk}
N.~Bobev, A.~M. Charles and V.~S. Min, \emph{{Euclidean Black Saddles and
  AdS$_4$ Black Holes}},  \href{http://arxiv.org/abs/2006.01148}{{\tt
  2006.01148}}.

\bibitem{Nishimura:2012jh}
M.~Nishimura and Y.~Tanii, \emph{{Coupling of the BLG theory to a conformal
  supergravity background}},
  \href{http://dx.doi.org/10.1007/JHEP01(2013)120}{\emph{JHEP} {\bf 01} (2013)
  120}, [\href{http://arxiv.org/abs/1206.5388}{{\tt 1206.5388}}].

\bibitem{Genolini:2016sxe}
P.~Benetti~Genolini, D.~Cassani, D.~Martelli and J.~Sparks, \emph{{The
  holographic supersymmetric Casimir energy}},
  \href{http://dx.doi.org/10.1103/PhysRevD.95.021902}{\emph{Phys. Rev. D} {\bf
  95} (2017) 021902}, [\href{http://arxiv.org/abs/1606.02724}{{\tt
  1606.02724}}].

\bibitem{Closset:2013vra}
C.~Closset, T.~T. Dumitrescu, G.~Festuccia and Z.~Komargodski, \emph{{The
  Geometry of Supersymmetric Partition Functions}},
  \href{http://dx.doi.org/10.1007/JHEP01(2014)124}{\emph{JHEP} {\bf 01} (2014)
  124}, [\href{http://arxiv.org/abs/1309.5876}{{\tt 1309.5876}}].

\bibitem{Closset:2019ucb}
C.~Closset, L.~Di~Pietro and H.~Kim, \emph{{'t Hooft anomalies and the
  holomorphy of supersymmetric partition functions}},
  \href{http://dx.doi.org/10.1007/JHEP08(2019)035}{\emph{JHEP} {\bf 08} (2019)
  035}, [\href{http://arxiv.org/abs/1905.05722}{{\tt 1905.05722}}].

\bibitem{Papadimitriou:2017kzw}
I.~Papadimitriou, \emph{{Supercurrent anomalies in 4d SCFTs}},
  \href{http://dx.doi.org/10.1007/JHEP07(2017)038}{\emph{JHEP} {\bf 07} (2017)
  038}, [\href{http://arxiv.org/abs/1703.04299}{{\tt 1703.04299}}].

\bibitem{An:2017ihs}
O.~S. An, \emph{{Anomaly-corrected supersymmetry algebra and supersymmetric
  holographic renormalization}},
  \href{http://dx.doi.org/10.1007/JHEP12(2017)107}{\emph{JHEP} {\bf 12} (2017)
  107}, [\href{http://arxiv.org/abs/1703.09607}{{\tt 1703.09607}}].

\bibitem{Kuzenko:2019vvi}
S.~M. Kuzenko, A.~Schwimmer and S.~Theisen, \emph{{Comments on Anomalies in
  Supersymmetric Theories}},
  \href{http://dx.doi.org/10.1088/1751-8121/ab64a8}{\emph{J. Phys. A} {\bf 53}
  (2020) 064003}, [\href{http://arxiv.org/abs/1909.07084}{{\tt 1909.07084}}].

\bibitem{Back:1978zf}
A.~Back, P.~G. Freund and M.~Forger, \emph{{New Gravitational Instantons and
  Universal Spin Structures}},
  \href{http://dx.doi.org/10.1016/0370-2693(78)90616-0}{\emph{Phys. Lett. B}
  {\bf 77} (1978) 181--184}.

\bibitem{Avis:1979de}
S.~Avis and C.~Isham, \emph{{Generalized spin structures on four dimensional
  space-times}}, \href{http://dx.doi.org/10.1007/BF01197630}{\emph{Commun.
  Math. Phys.} {\bf 72} (1980) 103}.

\bibitem{deWit:1982bul}
B.~de~Wit and H.~Nicolai, \emph{{N=8 Supergravity}},
  \href{http://dx.doi.org/10.1016/0550-3213(82)90120-1}{\emph{Nucl. Phys. B}
  {\bf 208} (1982) 323}.

\bibitem{CSF}
E.~Cremmer, J.~Scherk and S.~Ferrara, \emph{{SU(4) Invariant Supergravity
  Theory}}, \href{http://dx.doi.org/10.1016/0370-2693(78)90060-6}{\emph{Phys.
  Lett. B} {\bf 74} (1978) 61--64}.

\bibitem{Amsel:2008iz}
A.~J. Amsel and D.~Marolf, \emph{{Supersymmetric Multi-trace Boundary
  Conditions in AdS}},
  \href{http://dx.doi.org/10.1088/0264-9381/26/2/025010}{\emph{Class. Quant.
  Grav.} {\bf 26} (2009) 025010}, [\href{http://arxiv.org/abs/0808.2184}{{\tt
  0808.2184}}].

\end{thebibliography}\endgroup
}

\end{document}